\documentclass[twocolumn, twocolappendix]{aastex7}

\makeatletter
\def\@outputaffil@separated{}  % disables affils in main author list
\makeatother
\newcommand\minesweeper{\texttt{MINESweeper}}
\newcommand\minesweepercat{\texttt{BOSS-MINESweeper}}
\newcommand\kms{km\,s$^{-1}$}
\newcommand\lunit{kpc\,km\,s$^{-1}$}
\newcommand\segcat{\texttt{SEGUE-MINESweeper}}

\shorttitle{Mapping the Distant and Metal-Poor Milky Way with SDSS-V}
\shortauthors{Chandra et al.}

\begin{document}

\title{Mapping the Distant and Metal-Poor Milky Way with SDSS-V}

\correspondingauthor{Vedant Chandra}
\email{vedant.chandra@cfa.harvard.edu}

\author[0000-0002-0572-8012]{Vedant Chandra}
\affiliation{Center for Astrophysics $\mid$ Harvard \& Smithsonian, 60 Garden St, Cambridge, MA 02138, USA}
\email{vedant.chandra@cfa.harvard.edu}

\author[0000-0002-1617-8917]{Phillip~A.~Cargile}
\affiliation{Center for Astrophysics $\mid$ Harvard \& Smithsonian, 60 Garden St, Cambridge, MA 02138, USA}
\email{pcargile@cfa.harvard.edu}

\author[0000-0002-4863-8842]{Alexander~P.~Ji}
\affiliation{Department of Astronomy \& Astrophysics, University of Chicago, 5640 S Ellis Avenue, Chicago, IL 60637, USA}
\affiliation{Kavli Institute for Cosmological Physics, University of Chicago, Chicago, IL 60637, USA}
\affiliation{NSF-Simons AI Institute for the Sky (SkAI), 172 E. Chestnut St., Chicago, IL 60611, USA}
\email{alexji@uchicago.edu}

\author[0000-0002-1590-8551]{Charlie~Conroy}
\affiliation{Center for Astrophysics $\mid$ Harvard \& Smithsonian, 60 Garden St, Cambridge, MA 02138, USA}
\email{cconroy@cfa.harvard.edu}

\author[0000-0003-4996-9069]{Hans-Walter Rix}
\affiliation{Max-Planck-Institut f{\"u}r Astronomie, K{\"o}nigstuhl 17, D-69117 Heidelberg, Germany}
\email{rix@mpia.de}

\author[0000-0002-6993-0826]{Emily~Cunningham}
\affiliation{Department of Astronomy, Columbia University, 550 West 120th Street, New York, NY, 10027, USA}
\affiliation{Department of Astronomy, Boston University, 725 Commonwealth Ave., Boston, MA 02215, USA}
\email{ecc2198@columbia.edu}

\author[0000-0003-4254-7111]{Bruno Dias}
\affiliation{Instituto de Astrof{\'i}sica, Departamento de F{\'i}sica y Astronom{\'i}a, Facultad de Ciencias Exactas, Universidad Andres Bello, Santiago, Chile}
\email{astro.bdias@gmail.com}

\author[0000-0003-3922-7336]{Chervin Laporte}
\affiliation{LIRA, Observatoire de Paris, CNRS, 92190 Meudon, France}
\affiliation{Institut de Ci\`{e}ncies del Cosmos, Universitat de Barcelona, Mart\'{i} Franqu\`{e}s 1, 08028 Barcelona, Spain}
\affiliation{Kavli IPMU (WPI), UTIAS, The University of Tokyo, Kashiwa, Chiba 277-8583, Japan}
\email{chervin.laporte@icc.ub.edu}

\author[0000-0003-1697-7062]{William Cerny}
\affiliation{Department of Astronomy, Yale University, New Haven, CT 06520, USA}
\email{william.cerny@yale.edu}

\author[0000-0002-9269-8287]{Guilherme Limberg} 
\affiliation{Department of Astronomy \& Astrophysics, University of Chicago, 5640 S Ellis Avenue, Chicago, IL 60637, USA}
\affiliation{Kavli Institute for Cosmological Physics, University of Chicago, Chicago, IL 60637, USA}
\email{limberg@uchicago.edu}

%%%%%%%%%%%%%%%%%%%%%%%%%%%%

\author[0000-0002-8304-5444]{Avrajit Bandyopadhyay}
\affiliation{Department of Astronomy, University of Florida, Bryant Space Science Center, Stadium Road, Gainesville, FL 32611, USA}
\email{abandyopadhyay@ufl.edu}

\author[0000-0002-7846-9787]{Ana~Bonaca}
\affiliation{The Observatories of the Carnegie Institution for Science, 813 Santa Barbara Street, Pasadena, CA 91101, USA}
\email{abonaca@carnegiescience.edu}

\author[0000-0003-0174-0564]{Andrew R. Casey}
\affiliation{School of Physics \& Astronomy, Monash University, Wellington Road, Clayton, Victoria 3800, Australia}
\email{andrew.casey@monash.edu}

\author[0009-0000-4049-5851]{John Donor}
\affiliation{Department of Physics \& Astronomy, Texas Christian University, Fort Worth, TX 76129, USA}
\email{j.donor@tcu.edu}

\author[0000-0003-3526-5052]{Jos\'e G. Fern\'andez-Trincado}
\affiliation{Instituto de Astronom\'{i}a, Universidad Cat\'{o}lica del Norte, Av. Angamos 0610, Antofagasta, Chile}
\email{jose.fernandez@ucn.cl}

\author[0000-0002-0740-8346]{Peter M.~Frinchaboy} 
\affiliation{Department of Physics \& Astronomy, Texas Christian University, Fort Worth, TX 76129, USA}
\email{p.frinchaboy@tcu.edu}

\author[0000-0002-3956-2102]{Pramod Gupta}
\affiliation{Department of Astronomy, University of Washington, Box 351580, Seattle, WA 98195, USA}
\email{psgupta@uw.edu}

\author[0000-0002-1423-2174]{Keith Hawkins}
\affiliation{Department of Astronomy, University of Texas at Austin, Austin, TX 78712, USA}
\email{keithhawkins@utexas.edu}

\author[0000-0001-7258-1834]{Jennifer A. Johnson}
\affiliation{Department of Astronomy, The Ohio State University, 140 W.\,18th Ave., Columbus, OH 43210, USA}
\email{johnson.3064@osu.edu}

\author[0000-0001-9852-1610]{Juna A. Kollmeier}
\affiliation{The Observatories of the Carnegie Institution for Science, 813 Santa Barbara Street, Pasadena, CA 91101, USA}
\affiliation{Canadian Institute for Theoretical Astrophysics, University of Toronto, Toronto, ON M5S-98H, Canada}
\affiliation{Canadian Institute for Advanced Research, 661 University Avenue, Suite 505, Toronto, ON M5G 1M1 Canada}
\email{jak@carnegiescience.edu}

\author[0000-0001-7297-8508]{Madeline Lucey}
\affiliation{Department of Physics and Astronomy, University of Pennsylvania, Philadelphia, PA 19104, USA}
\email{m_lucey@utexas.edu}

\author[0000-0003-3410-5794]{Ilija Medan}
\affiliation{Department of Physics and Astronomy, Vanderbilt University, 6301 Stevenson Center Ln., Nashville, TN 37235, USA}
\email{ilija.medan@vanderbilt.edu}

\author[0000-0001-8237-5209]{Szabolcs M\'esz\'aros}
\affiliation{ELTE Gothard Astrophysical Observatory, H-9704 Szombathely, Szent Imre herceg st. 112, Hungary}
\affiliation{MTA-ELTE Lend{\"u}let ``Momentum'' Milky Way Research Group, Hungary}
\email{meszi@gothard.hu}

\author[0000-0002-6770-2627]{Sean Morrison}
\affiliation{Department of Astronomy, University of Illinois at Urbana-Champaign, Urbana, IL 61801, USA}
\email{smorris0@illinois.edu}

\author[0000-0003-2486-3858]{Jos\'e S\'anchez-Gallego}
\affiliation{Department of Astronomy, University of Washington, Box 351580, Seattle, WA 98195, USA}
\email{gallegoj@uw.edu}

\author[orcid=0000-0002-6561-9002,gname=Andrew,sname=Saydjari]{Andrew~K.~Saydjari} 
\altaffiliation{Hubble Fellow} 
\affiliation{Department of Astrophysical Sciences, Princeton University, Princeton, NJ 08544 USA} \email{aksaydjari@gmail.com}

\author[0000-0002-4454-1920]{Conor Sayres}
\affiliation{Department of Astronomy, University of Washington, Box 351580, Seattle, WA 98195, USA}
\email{csayres@uw.edu}

\author[0000-0001-5761-6779]{Kevin C.\ Schlaufman}
\affiliation{William H.\ Miller III Department of Physics \& Astronomy, Johns Hopkins University, 3400 N Charles St, Baltimore, MD 21218, USA}
\email{kschlaufman@jhu.edu}

\author[0000-0002-3481-9052]{Keivan G. Stassun}
\affiliation{Department of Physics and Astronomy, Vanderbilt University, 6301 Stevenson Center Ln., Nashville, TN 37235, USA}
\email{keivan.stassun@vanderbilt.edu}

\author[0000-0002-4818-7885]{Jamie Tayar}
\affiliation{Department of Astronomy, University of Florida, Bryant Space Science Center, Stadium Road, Gainesville, FL 32611, USA}
\email{jtayar@ufl.edu}

\author[0000-0003-0179-9662]{Zachary Way}
\affiliation{Department of Physics and Astronomy, Georgia State University, Atlanta, GA 30302, USA}
\email{zway1@gsu.edu}

\begin{abstract}
\noindent
The fifth-generation Sloan Digital Sky Survey (SDSS-V) is conducting the first all--sky low--resolution spectroscopic survey of the Milky Way's stellar halo. 
We describe the stellar parameter pipeline for the SDSS-V halo survey, which simultaneously models spectra, broadband photometry, and parallaxes to derive stellar parameters, metallicities, alpha abundances, and distances. 
The resulting \minesweepercat{} catalog is validated across a wide range of stellar parameters and metallicities using star clusters and a comparison to high-resolution spectroscopic surveys.
We demonstrate several scientific capabilities of this dataset: identifying the most chemically peculiar stars in our Galaxy, discovering and mapping distant halo substructures, and measuring the all--sky dynamics of the Milky Way on the largest scales. 
The \minesweepercat{} catalog for SDSS DR19 is publicly available and will be updated for future data releases. 
\end{abstract}


\section{Introduction} \label{sec:intro}

The stellar halo of the Milky Way (MW) encodes a record of our galaxy's merger history and chemodynamical evolution \citep[e.g.,][]{Bullock2005, Helmi2008, Belokurov2013b, Helmi2020}. 
The practice of reconstructing the MW's history through these stellar relics has been termed `galactic archaeology' \citep[e.g.,][]{Freeman2002, Helmi2020, Deason2024}. 
In recent years, this field has been accelerated by
the advent of wide-field spectroscopic surveys that deliver radial velocities and chemical abundances for millions of stars \citep[e.g.,][]{Yanny2009, Majewski2017, Katz2022, Cooper2023}. 
In concert with astrometry --- parallaxes and proper motions --- from the \textit{Gaia} space observatory \citep{GaiaCollaboration2016, GaiaCollaboration2018, GaiaCollaboration2021}, the full 6D dynamics of halo populations can be measured, revealing their progenitors \citep[e.g.,][]{Helmi2018, Belokurov2018a, Naidu2020, Deason2024}. 

The stellar halo hosts two exciting frontiers of discovery: the most distant, and the most metal-poor stars. 
Stars in the most distant outer halo have the longest orbital timescales, and consequently retain a strong spatial and dynamical memory of past mergers \citep[e.g.,][]{Hendel2015, Fattahi2020}. 
A large number of accreted dwarf galaxies (hereafter `dwarfs') are expected to populate the outer halo, sometimes leaving coherent imprints like shells and streams \citep[e.g.,][]{Newberg2002, Majewski2003, Grillmair2006c, Belokurov2007a, Deason2018, Simon2019, Helmi2020, Chandra2023a, Chandra2023b}. 
Additionally, stars in the outer halo can be used to measure the global dynamics of the MW on the largest scales, which are dominated by the influence of the massive infalling Large Magellanic Cloud \citep[e.g.,][]{Erkal2019a, Erkal2020, Erkal2021, Petersen2020, Petersen2021, Yaaqib2024, Chandra2025}. 

The early history of the MW is also encoded in the chemical fingerprints of the Galaxy's most metal-poor stars. 
These stars --- with total metal content below $\lesssim 1\%$ the solar level --- were formed out of gas enriched by the first massive stars in the Galaxy \citep[e.g.,][]{Bromm2004}. 
Their chemical makeup therefore provides a window into the physical processes of stellar birth and death in the nascent MW \citep[e.g.,][]{Umeda2002, Cayrel2004, Meynet2006, Heger2010, Caffau2011, Ezzeddine2019, Skuladottir2021, Yong2021, Bonifacio2025}. 
Past searches --- including with previous generations of SDSS --- have yielded hundreds of stars with [Fe/H]$\lesssim -3.0$ \citep[e.g.][]{Beers1992, Cayrel2004, Frebel2006, Schlaufman2014, Aguado2016, Starkenburg2017, Li2018, Placco2025}.

Across both these frontiers, progress has accelerated due to the advent of new observational facilities. 
\textit{Gaia} parallaxes can be used to effectively filter out foreground dwarf stars, isolating distant giants in the outer halo. 
Furthermore, low-resolution \textit{Gaia} `XP' spectra have recently delivered metallicity estimates for over a hundred million stars across the Galaxy, yielding thousands of promising metal-poor candidates \citep[][Thai et al, in prep]{Andrae2023, Zhang2023, Yao2024}. 
The ongoing `MAGIC' survey is utilizing narrow-band imaging on the Dark Energy Camera (DECam; \citealt{Flaugher2015}) to identify metal-poor stars at larger distances than ever before \citep{Placco2025}. 
All of these distant and metal-poor candidates require spectroscopic observations to measure their distances and metallicities, and unlock their scientific potential through their kinematics and full abundance patterns.

The fifth-generation Sloan Digital Sky Survey (SDSS-V; \citealt{Kollmeier2025}) is the most comprehensive all-sky spectroscopic survey to date, covering both hemispheres with optical (BOSS; \citealt{Smee2013}) and infrared (APOGEE; \citealt{Wilson2019apogee}) spectrographs. 
SDSS-V consists of three `mappers': Milky Way Mapper (MWM), Black Hole Mapper (BHM), and Local Volume Mapper (LVM; \citealt{Drory2024}). 
By the end of the survey, MWM is expected to observe over ten million stars in the optical and infrared. 
The vast majority of these stars will reside within the disk of the MW.
However, within MWM, the halo survey specifically targets distant and metal-poor stars using a variety of techniques. 
These stars are given a high priority during target allocation, ensuring that a large number of halo stars will be observed over the course of the survey.
A large majority of halo stars are too faint to be observed with APOGEE, and consequently only have BOSS spectra.
To date, over a hundred thousand stars have been observed with BOSS as a part of the MWM halo program, out of half a million expected by the end of the survey.

Turning spectra into scientifically-useful catalogs requires robust pipelines to derive stellar parameters from spectroscopic observations --- temperatures, surface gravities, metallicities, and chemical abundances. 
For kinematic analyses of the halo, isochrone-based distances are especially important, since \textit{Gaia} parallaxes are not precise enough to deliver reliable distances beyond $\gtrsim 10$~kpc for most stars. 
Several automated routines produce stellar parameters for SDSS-V BOSS spectra, broadly packaged into the \texttt{astra} framework \citep[][Casey et al., in prep]{Sizemore2024}. 
However, while these routines are excellent for the vast majority of stars, they can struggle to produce reliable parameters for rare halo stars. 
For example, pipelines that only use spectroscopic information struggle to differentiate hot stars from metal-poor stars, since both have weak absorption lines.
However, this degeneracy can be broken with photometric information, which provides an independent measurement of the temperature. 
It is therefore desirable to develop a dedicated inference pipeline for the MWM halo program that models all available spectro-photometry and samples the entire parameter space.

In this paper, we describe the \minesweepercat{} value-added catalog from the SDSS-V halo survey, which derives stellar parameters from BOSS spectra using the \minesweeper{} fitting routine \citep{Cargile2020}. 
\minesweeper{} incorporates all available information about a star --- the BOSS spectrum, archival broadband photometry, and the \textit{Gaia} parallax --- and determines the best-fitting stellar parameters with a full Bayesian fitting scheme. 
\minesweeper{} simultaneously fits isochrone models to the star, delivering evolutionary parameters and spectrophotometric distances. 

\setcounter{footnote}{0}

In $\S$\ref{sec:catalog}, we briefly describe the SDSS-V halo targeting and data, and provide a detailed description of the \minesweeper{} code and \minesweepercat{} catalog. 
A validation of our \minesweepercat{} catalog is presented in $\S$\ref{sec:validation}, using dedicated observations of star clusters and a comparison to the the high-resolution APOGEE survey. 
We showcase the scientific capabilities of \minesweepercat{} in $\S$\ref{sec:science}, from identifying chemically rare stars to mapping all-sky dynamics of the Galactic outskirts. 
Finally, we contextualize our results and conclude in $\S$\ref{sec:discuss}, providing links to the latest public release of the catalog. 

During the development of our pipeline, we fit $\approx~100,000$ archival spectra from the Sloan Extension for Galactic Understanding and Exploration (SEGUE; \citealt{Yanny2009}) --- this catalog is described and released in Appendix~\ref{sec:segcat}. 
The star cluster validation is tabulated in Appendix~\ref{sec:clustertab}, and a full datamodel of the \minesweepercat{} catalog is presented in Appendix~\ref{sec:datamodel}.

\section{Data \& Methods}\label{sec:catalog}

\subsection{BOSS Spectra and Halo Target Selection}\label{sec:spectra}

The Baryon Oscillation Spectroscopic Survey (BOSS; \citealt{Smee2013}) spectrograph delivers $R \approx 1800$ spectra across the full optical range from $3650-9500\,$\AA.
One BOSS spectrograph remains mounted to the 2.5\,m telescope at Apache Point Observatory (APO), and the other spectrograph has been moved to the 2.5\,m DuPont telescope \citep{Bowen1973} at Las Campanas Observatory (LCO), providing all-sky coverage. 
Prior to 2021, SDSS-V continued using the plug-plate fiber positioner from past SDSS iterations. 
In 2021, a new robotic fiber positioner was commissioned, enabling more flexibility and faster re-configurations between fields \citep{Pogge2020}. 
All the BOSS data analyzed in this work were reduced using \texttt{IDLspec2D v6\_2\_0} (\citealt{Bolton2012, Dawson2013}; Morrison et al, in prep).
Note that the publicly released spectra in DR19 use an older version, \texttt{v6\_1\_3}. 
Future data releases will use updated versions of the pipeline, which is actively under development. 

The vast majority of the stars observed by MWM reside in the disk of the MW. 
Stars in the stellar halo are specifically targeted by a number of `cartons' or selection categories. 
The highest-priority halo cartons are pure selections of the most distant and the most metal-poor candidates. 
Distant red giant branch stars are selected following the methodology outlined in \citet[][see their $\S$2.1]{Chandra2023b}. 
Briefly, plausible giants are isolated from far more numerous nearby dwarfs, on the basis of \textit{Gaia} DR3 parallaxes. 
The sample is further purified with a selection based on infrared colors from the WISE space observatory \citep{Mainzer2014}, and approximate photometric distances are calculated using a [Fe/H]$=-1.2$, 10~Gyr MIST isochrone \citep{Choi2016}. 
A similar selection is described in \cite{Conroy2018, Conroy2021}. 
This selection has produced a pure sample of red giants out to 100~kpc and beyond \citep[e.g.,][]{Chandra2025}. 
Giants with estimated distances $\gtrsim 30$~kpc are given the highest priority for SDSS-V targeting, followed by those with estimated distances  $\gtrsim 10$~kpc. 

Metal-poor stars are selected based on metallicities measured from low-resolution \textit{Gaia} `XP' prism spectroscopy. 
Specifically, we use the catalog of \cite{Andrae2023}, who measured XP-based metallicities for 175~million stars. 
We adopt several quality cuts to ensure robust metallicities: $\log{g} < 4.0$, $T_\mathrm{eff} < 5500$\,K, $M_\mathrm{W1} > -0.3 - 0.006 \times (5500 - T_\mathrm{eff})$, and $M_\mathrm{W1} > -0.01 \times (5300 - T_\mathrm{eff})$, where $M_\mathrm{W1} \equiv \mathrm{W1} + 5 \log_{10}{(\varpi/100)}$ is the absolute magnitude in WISE. 
The latter two cuts are primarily designed to remove reddened hot (OBA-type) stars from the sample, since they have XP spectra resembling cooler metal-poor stars. 
Stars with XP-inferred metallicity [Fe/H]$_\mathrm{XP} < -2.0$ are given the highest targeting priority, followed by stars with [Fe/H]$_\mathrm{XP} < -1.5$ and finally [Fe/H]$_\mathrm{XP} < -1.0$. 

The \minesweepercat{} catalog also includes a variety of other selection cartons that have a high fraction of halo stars. 
This includes metal-poor stars selected using an IR-based selection from the 2MASS photometric survey \citep{Schlaufman2014}, and metal-poor stars selected using narrow-band photometry from the SkyMapper survey \citep{Wolf2018}. 

One important MWM halo target class that is \textit{omitted} from the \minesweepercat{} catalog is the `local halo' carton. 
These are fast-moving K and M dwarf stars that predominantly reside in the solar neighborhood. 
Due to systematic uncertainties in isochrones for such cool dwarfs, \minesweeper{} struggles to produce reliable parameters for these stars (as do many other pipelines). 
Therefore, we have focused the \minesweepercat{} catalog towards turnoff and giant stars in the halo, and defer to other catalogs \citep[e.g.,][]{Sizemore2024} for cool dwarfs. 

The LCO footprint includes the Large and Small Magellanic Clouds (hereafter Clouds), which are targeted as a part of the Magellanic Genesis program in SDSS-V. 
Stars targeted as belonging to the Clouds are excluded from the \minesweepercat{} catalog, since they require a set of priors that are markedly different from field MW halo stars. 
We are developing a dedicated \minesweeper{} pipeline for stars in the Clouds, which may be released with future SDSS-V data releases that include LCO data. 
Finally, stars with Galactic latitude $|b| < 5^\circ$ are excluded, because it is challenging to uniquely identify associated photometry for stars in these crowded regions, and these fields are dominated by stars in the MW disk. 

\begin{figure}
    \centering
    \includegraphics[width=\columnwidth]{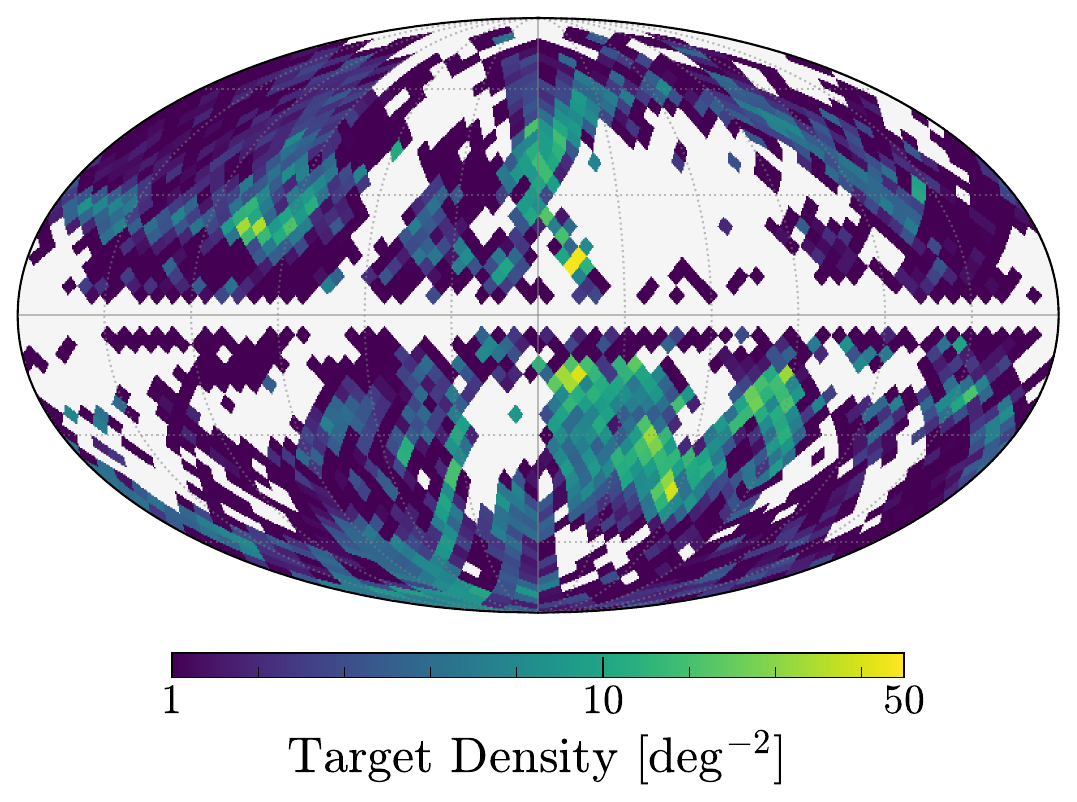}
    \includegraphics[width=\columnwidth]{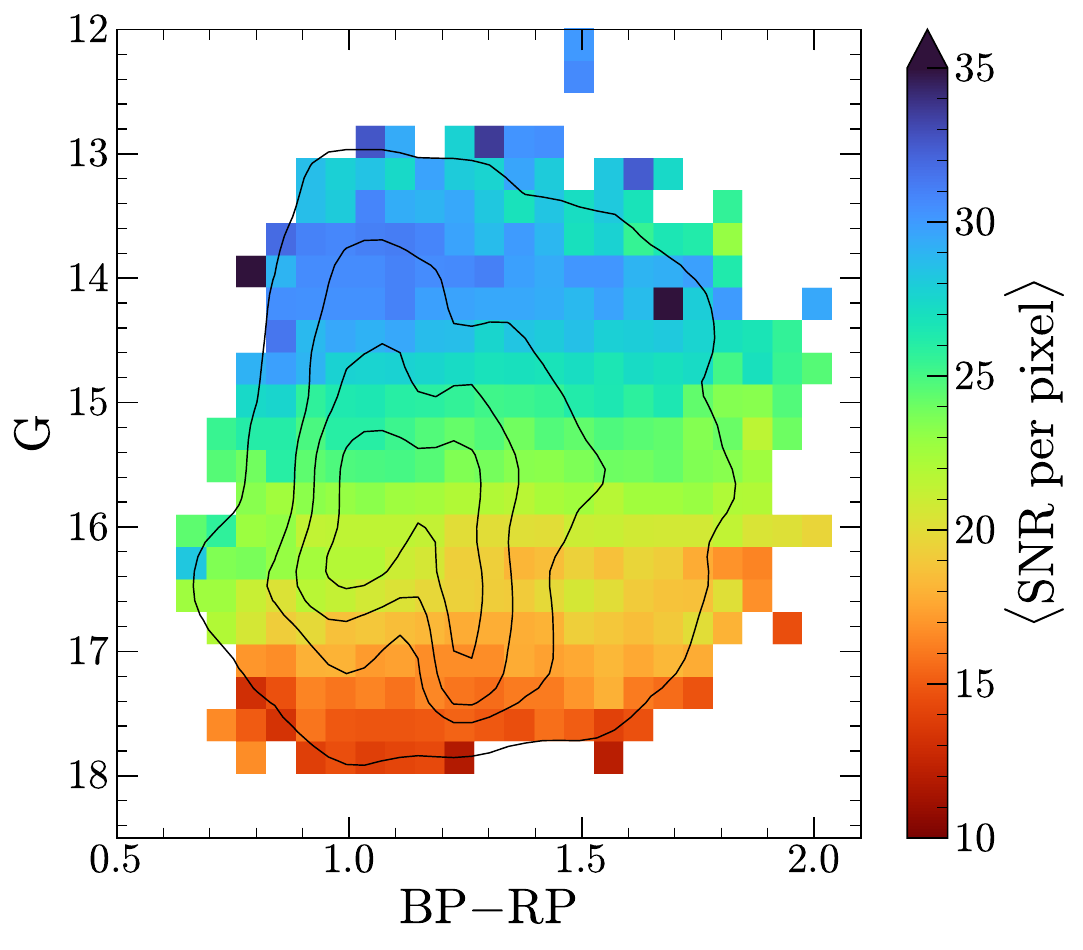}
    \caption{\textbf{Top:} Distribution of SDSS-V halo stars observed up through 2024, in Galactic coordinates. 
    This figure includes data observed from both APO and LCO, although only APO data are released in DR19. 
    \textbf{Bottom:} Stellar distribution in \textit{Gaia} color-magnitude space, colored by median signal-to-noise ratio (in the $4750-5500$\,\AA{} region) of the co-added BOSS spectrum. 
    Contours of target density are overlaid.}
    \label{fig:targets-sky}
\end{figure}

SDSS-V halo targets observed to date are shown in Figure~\ref{fig:targets-sky}. 
The top panel shows the target density on-sky, and the bottom panel shows the color-magnitude space colored by the median SNR of the BOSS spectrum. 
A key feature of SDSS-V is continuous coverage across both hemispheres, making it the first all-sky spectroscopic survey of the MW halo. 
Several large gaps remain in the present data, which will be filled in by the end of the survey.
It is a science requirement of the halo survey that any remaining gaps be smaller than $7~\mathrm{deg^{-2}}$. 
This ensures that the field halo is well-sampled spatially, enabling measurements of the global dynamics of the distant MW halo \citep[e.g.,][]{Chandra2025}. 

The majority of stars observed by the MWM halo program only have a single BOSS exposure. 
However, around $15\%$ of halo targets received multiple exposures, sometimes split over several days. 
The BOSS pipeline coadds exposures observed within a given night. 
For each unique star (i.e., a unique \textit{Gaia} source ID), we further co-add all available SDSS-V spectra when available, using the \texttt{coadd1d} module from \texttt{PypeIt} \citep{pypeit:joss_pub, pypeit:zenodo}. 
Note that there is no attempt to remove radial velocity variations between exposures --- for example due to binary motion --- before either co-adding step. 

Although the scientific results in this paper utilize all-sky SDSS-V data from APO and LCO, only APO data observed before MJD~60130 (2023~07~05) are publicly available as a part of DR19, and consequently only parameters for those stars are currently released \citep{SDSSCollaboration2025}. 
Future data releases will include \minesweepercat{} parameters for APO and LCO stars. 

\subsection{Stellar Parameters with \minesweeper{}}\label{sec:mscode}

\begin{figure*}
    \centering
    \includegraphics[width=\textwidth]{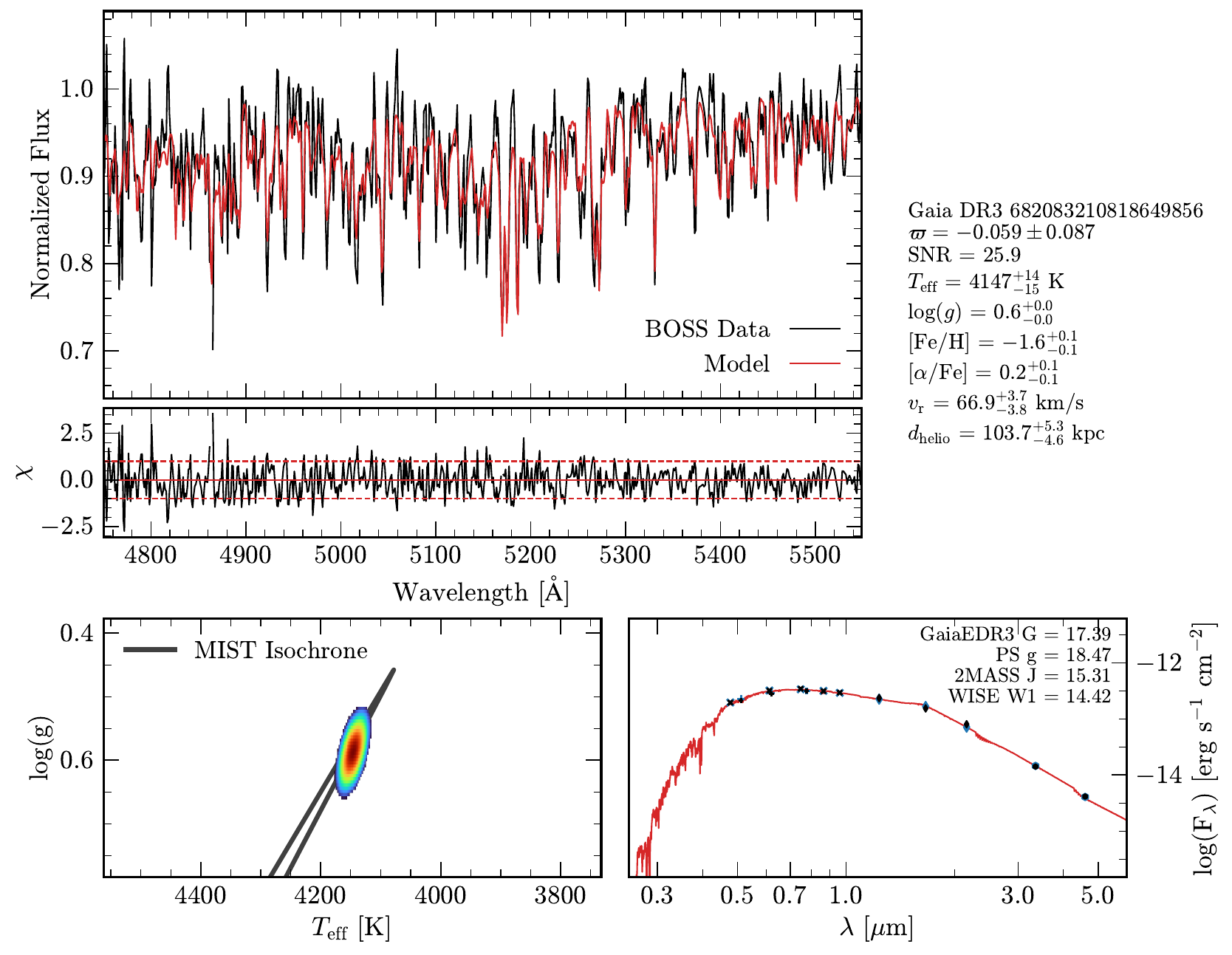}
    \caption{An example of a \minesweeper{} fit to a distant giant star ($d_\mathrm{helio} > 100$~kpc) from SDSS-V, demonstrating all of the information that is utilized. 
    \textbf{Top Left:} The fitted region of the SDSS-V spectrum (black), with the best-fitting model spectrum (red). The lower panel shows the error-scaled residual, with dashed lines indicating $\pm 1\sigma$ deviations. 
    \textbf{Top Right:} Best-fitting stellar parameters with $1\sigma$ uncertainties derived from the posterior distributions. The \textit{Gaia} parallax is also indicated, and is incorporated in the \minesweeper{} likelihood. 
    \textbf{Bottom Left:} Kiel diagram showing the posterior contours of $T_\mathrm{eff}$ and $\log{g}$. The black line shows the best-fitting MIST isochrone, from which parameters like the stellar age and distance are calculated. 
    \textbf{Bottom Right:} Broadband photometry of the star, with the best-fitting model spectral energy distribution overlaid in red. 
    } 
    \label{fig:fit} 
\end{figure*}

We estimate stellar parameters using the  \minesweeper{} code \citep[][]{Cargile2020}. 
\minesweeper{} is a Bayesian software routine that simultaneously fits spectroscopic, photometric, and parallax data to find the best-fitting model for a given star (see Figure~\ref{fig:fit} for a demonstration).
The inclusion of photometry is a key feature of \minesweeper{} that is particularly useful to break the spectroscopic degeneracies between temperature, surface gravity, and metallicity. 
During the \minesweeper{} fit, solutions are constrained to follow isochrones from MIST \citep{Dotter2016, Choi2016}.
This restricts solutions to a physically-plausible parameter space, and additionally enables us to apply priors on evolutionary parameters like stellar mass and age.  
As detailed in the previous section, \minesweeper{} is run once per \textit{Gaia} source, with all SDSS-V spectra for that source being co-added. 
Our main adopted priors are listed in Table~\ref{tab:priors}, and we defer to \cite{Cargile2020} for a more detailed discussion. 

\begin{deluxetable}{cc}
\label{tab:priors}
\tablecaption{Adopted priors in \minesweeper{}.}
\tablehead{\colhead{\hspace{1cm}Parameter}\hspace{1cm} & \colhead{Prior}\hspace{1cm}}
\tablewidth{\columnwidth}
\startdata
$T_\mathrm{eff}$ & $\mathcal{U}(3500, 9000)$ K \\
$\log{g}$ & $\mathcal{U}(0.0, 5.0)$ \\
$\mathrm{[Fe/H]}$ & $\mathcal{U}(-4.0, +0.5)$ \\
$\mathrm{[\alpha/Fe]}$ & $\mathcal{U}(-0.2, +0.6)$ \\
$V_\mathrm{\star} \sin{i}$ & $\mathcal{B}(1.05, 1.5) \ast \mathcal{U}(0, 250)$ \\
$M_{\ast, \mathrm{initial}}$ & \cite{Kroupa2001} IMF \\
EEP\tablenotemark{a} & $\mathcal{U}(100, 808)$ \\
Age & Galactic Age Prior\tablenotemark{b} \\
Distance & Galactic Density Prior\tablenotemark{b} \\
$A_\mathrm{V}$ & SFD\tablenotemark{c} Dustmap $\pm~20\%$ \\
\enddata
\tablenotetext{a}{Equivalent evolutionary phase on the MIST isochrone, see \cite{Dotter2016}}
\tablenotetext{b}{See \cite{Cargile2020} for details}
\tablenotetext{c}{\cite{Schlegel1998}, re-normalized by \cite{Schlafly2011}}
\end{deluxetable} %

\minesweeper{} compares the BOSS spectrum and broadband photometry to synthetic model spectra.
These spectra are calculated in 1D LTE using the radiative transfer codes \texttt{ATLAS12} and \texttt{SYNTHE} \citep{Kurucz1970, Kurucz1981, Kurucz1993}. 
The line list has been empirically tuned to match high-resolution spectra of the Sun and Arcturus. 
A constant microturbulence of $1$~\kms{} is assumed. 
We only fit the region between $4750-5550$\,\AA, since this is the region in which our linelists have been empirically calibrated (Figure~\ref{fig:fit}, top). 
This region includes the Balmer H$\beta$ absorption line and the \ion{Mg}{1}\,b triplet, in addition to several strong iron lines. 
To capture deviations between the relative flux calibration of the model and observed spectra, the model is scaled by a 4th-order Chebyshev polynomial during the fit, adding four free parameters. 

The synthetic spectrum grid is interpolated using fast neural network models, similar to those used in the \texttt{Payne} \citep{Ting2019}. 
Model spectra are calculated on a higher-resolution $R = 5000$ grid and convolved down to the $R \approx 1800$ BOSS resolution during the fitting process. 
We use the wavelength-dependent line spread function (LSF, assumed Gaussian) provided by the BOSS pipeline for this convolution, which is estimated for each spectrum using the widths of sky emission lines. 
Matching the instrumental resolution exactly would drive the stellar rotation parameter --- which is unresolved for most giants at BOSS resolution --- towards zero, leading to inefficient sampling during the fit. 
Furthermore, the empirically-determined LSF can be erroneous for some spectra. 
Therefore, we artificially sharpen the assumed LSF by 75\%,  and allow the free stellar rotation parameter to scale the LSF during the fit. 
Consequently, the rotation values provided in the catalog should not be physically interpreted, but can be diagnostically useful to catch poor-quality fits. 

We utilize archival optical--infrared broadband photometry in the likelihood (Figure~\ref{fig:fit}, bottom), drawing on photometry from \textit{Gaia}, SDSS, Pan-STARRS, 2MASS, and WISE, wherever available \citep{Fukugita1996, Gunn1998, Skrutskie2006,  Mainzer2014, Chambers2016, GaiaCollaboration2021, GaiaCollaboration2022}. 
A prior is placed on the extinction $A_\mathrm{V}$ following the \cite{Schlegel1998} dustmap (renormalized by \citealt{Schlafly2011}), with a $20\%$ uncertainty. 
The likelihood also uses \textit{Gaia}~DR3 parallax measurements, which have been zeropoint-corrected following the recipes described in \cite{Lindegren2021c}. 

The posterior distribution of stellar parameters is sampled with \texttt{dynesty} \citep{Speagle2020}, producing measurements of the radial velocity $v_\mathrm{r}$, effective temperature $T_\mathrm{eff}$, surface gravity $\log{g}$, metallicity [Fe/H], [$\alpha$/Fe] abundance, and heliocentric distance. 
The median of the distribution is reported as the best-fit value, and the difference between the 84th and 16th quantiles is reported as the $1\sigma$ uncertainty. 
The full data model for the catalog is presented in Appendix~\ref{sec:datamodel}. 
Although \minesweeper{} returns a wealth of information, some parameters should be treated with caution before being used in scientific analyses. 
For example, although a stellar age is returned for each star, these values are typically prior-dominated for most giant stars \cite[see][for a discussion on how to select reliable stellar ages]{Woody2025}. 
Therefore, the catalog ages should be treated with caution --- although they are useful to derive stellar parameters, it is quite challenging to interpret them in astrophysical terms \citep[e.g.,][]{Ying2023, Ying2025, Morales2025}.

Once the stellar parameters have been determined, dynamical quantities for each star are derived using \texttt{astropy} \citep{AstropyCollaboration2013, AstropyCollaboration2018, AstropyCollaboration2022} and \texttt{gala} \citep{gala,adrian_price_whelan_2020_4159870}. 
A right-handed Galactocentric frame is assumed with a solar position $\mathbf{x}_\odot = (-8.12, 0.00, 0.02)$~kpc, and solar velocity $\mathbf{v}_\odot = (12.9, 245.6, 7.8)$~$\mathrm{km\,s^{-1}}$ \citep{Reid2004,Drimmel2018,GravityCollaboration2019}. 
We adopt the \texttt{MilkyWayPotential2022} from \texttt{gala} when calculating orbital quantities like the total energy and eccentricity, which has a circular velocity at the solar position $v_\mathrm{c}(R_\odot) = 229~\textrm{km}~\textrm{s}^{-1}$ \citep{Eilers2019}. 
Orbits are calculated over $2.5$~Gyr with $1$~Myr timesteps. 

Although \minesweeper{} delivers robust stellar parameters for stars across the halo survey, it is significantly more resource-intensive than most spectroscopic inference pipelines. 
This is primarily because of the full posterior sampling in a Bayesian framework, which is crucial to find the correct solution and deliver well-calibrated parameter uncertainties. 
Several steps have been taken to make this sampling as fast as possible, for example by emulating the synthetic spectrum generation and isochrone interpolation using neural networks \citep[see][]{Ting2019, Cargile2020}. 
Regardless, we only run \minesweeper{} on the high-value subset of SDSS-V halo stars that were targeted to be metal poor or distant. 
\minesweeper{} is run locally at the FASRC~Cannon cluster operated by Harvard University. 
The pipeline takes $20-30$ core-minutes per star, including post-processing. 
Work is underway to further speed up the pipeline, for example by replacing the nested sampling with stochastic variational inference \citep[SVI;][]{Hoffman2013}. 

During the development of the \minesweepercat{} catalog, we also ran \minesweeper{} on archival spectra from the Sloan Extension for Galactic Understanding and Exploration (SEGUE; \citealt{Yanny2009, Abazajian2009}) survey. 
The resulting \segcat{} catalog of $\approx 100,000$ stars is described and publicly released in Appendix~\ref{sec:segcat} \citep{Cargile2025}. 

\subsection{Cleanliness Cuts}

A key goal of the \minesweepercat{} catalog is to provide a reliable, pure sample of stellar parameters that can be trusted for individual stars. 
Therefore, we apply several cleanliness cuts to the catalog to ensure that the reported stellar parameters are robust. 
These cuts have evolved over time, building on experience from the H3 Survey \citep{Conroy2019b}, which also uses the \minesweeper{} code over a similar wavelength regime. 

The following cuts are utilized to select stars with reliable parameters:

\begin{itemize}
    \item Median SNR $> 10$ per pixel in the co-added spectrum across the $4750-5500$\,\AA{} range.

    \item $T_\mathrm{eff} < 7000$~K: hotter stars have poorly-modeled isochrones --- for example on the blue horizontal branch --- and unreliable metallicities due to their weaker absorption lines.

    \item $\log{g} < 4.5$: the coolest dwarfs have poorly-modeled isochrones --- particularly their radii --- and unreliable metallicities due to molecular features.
    
    \item $V_\mathrm{rot} < 200$~\kms{}: removes unphysically large rotation velocities, which signal an unusual spectrum or poor data reduction.
    
    \item At least 3 photometric bands must be present (i.e., the three \textit{Gaia} bands at minimum). Most stars have $\gtrsim 10$ photometric measurements from various surveys.
    
    \item Reduced $\chi^2$ (spectroscopic $\chi^2$ divided by number of fitted pixels) less than $2$, which removes poor model fits and spurious spectra.
    
\end{itemize}

Out of all the BOSS spectra fit by \minesweeper{}, these cuts remove $\approx 20\%$ of stars. 
The publicly-released catalogs only contain stars that pass these reliability cuts. 

\section{Validating \minesweepercat{} Parameters}\label{sec:validation}

\subsection{Star Clusters}\label{sec:validation-clusters}

\begin{figure}
    \centering
    \includegraphics[width=0.97\columnwidth]{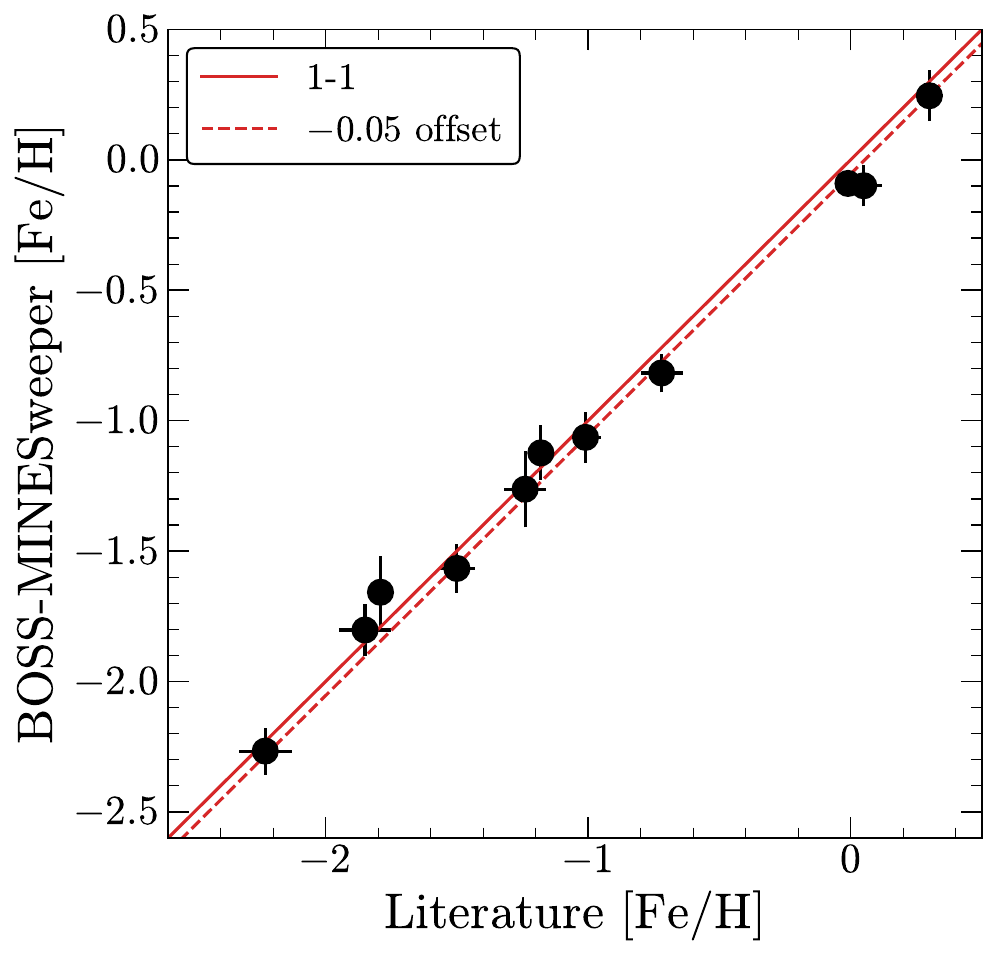}
    \includegraphics[width=\columnwidth]{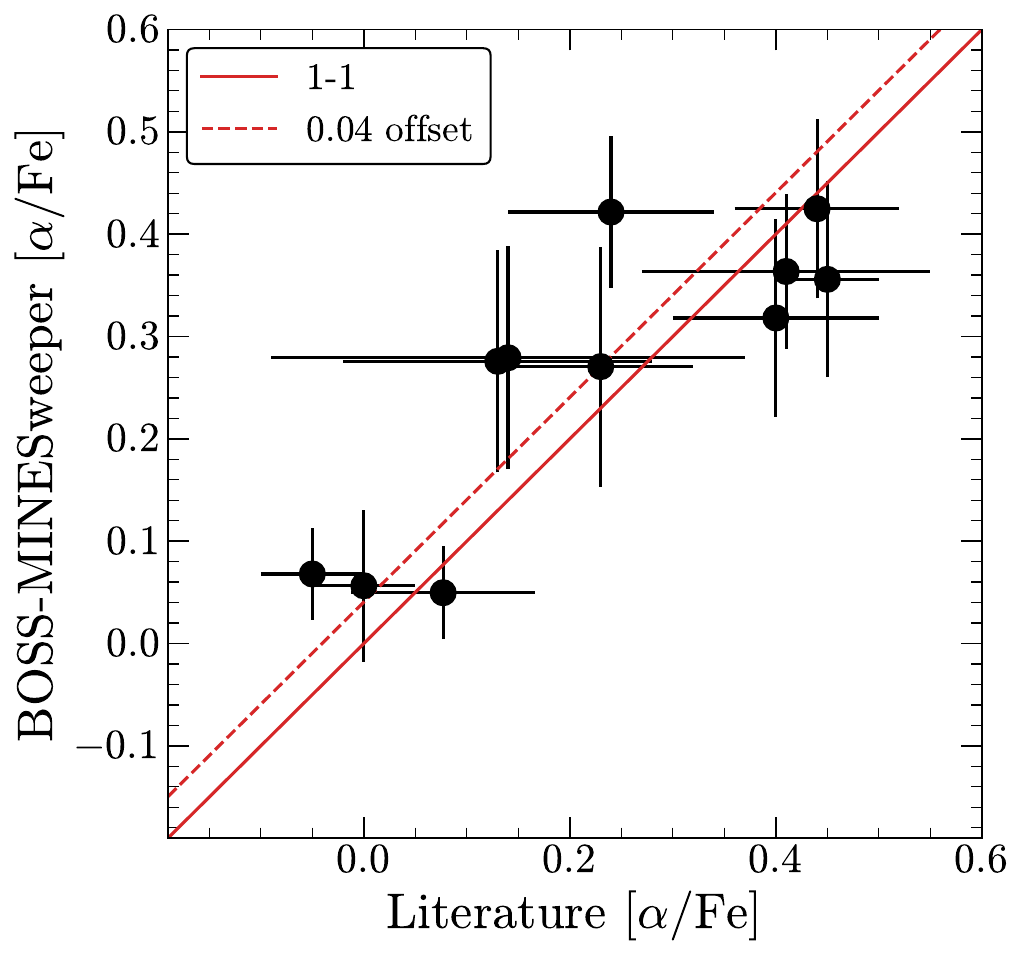}
    \caption{Mean metallicity and alpha abundance for calibration clusters in SDSS-V, as compared to parameters from the literature. 
    The underlying data and references are listed in Table~\ref{tab:clusters}.
    The 1-1 line is overlaid, along with the constant offset measured as the median difference between \minesweeper{} and literature. 
    }
    \label{fig:cluster_means}
\end{figure}

Star clusters have long been a gold standard to calibrate stellar parameters. 
As some of the earliest-known and most well-studied astronomical objects, star clusters have well-measured parameters and distances. 
To first order, stars in a cluster are assumed to have formed from the same birth material, and consequently have relatively homogeneous metallicities \cite[e.g.,][]{Gratton2004, Bastian2018, Krumholz2019}.
Consequently, clusters can be used to investigate whether spectroscopically-measured metallicities are systematically biased with respect to temperature and surface gravity. 

One important effect to consider is the influence of convective diffusion on the surface abundance of stars. 
Particularly for main sequence turnoff stars, the surface abundance of iron can be depleted by up to $0.2-0.3$~dex by diffusion, compared to the initial abundance of the star \citep{Dotter2017}. 
These effects are self-consistently incorporated in the \minesweeper{} fitting routine: the underlying MIST isochrones track the initial abundance and the evolutionary phase-dependent abundance. 
For a given star, the present-day surface metallicity is given by the \texttt{FeH} column, and the initial abundance from the isochrone model is given by \texttt{init\_FeH} (likewise for \texttt{aFe}). 
The choice of column depends on the science case; for example, the initial abundance is the relevant quantity when comparing to chemical evolution models, whereas the current abundance is relevant when comparing to other spectroscopic measurements of the surface abundance. 
For the cluster calibration in this section, we use the initial abundances reported by \minesweeper{}, which should reflect the initial chemical homogeneity of the cluster.

Figure~\ref{fig:cluster_means} compares the mean \minesweepercat{} metallicity and alpha abundance of calibration clusters as compared to established values from the literature. 
The literature values are drawn from a variety of sources, and are consistent with the homogeneous metallicity scale presented in \cite{Dias2016}. 
These results are also listed in Table~\ref{tab:clusters} in Appendix~\ref{sec:clustertab}, along with references to the cluster parameters. 
Since the \minesweepercat{} wavelength range is most sensitive to the \ion{Mg}{1}\,b triplet feature, we use literature Mg abundances as our point of comparison for the alpha abundances. 
We use the median difference in mean parameters across all clusters to define an offset between our abundance scale and the literature scale (which is predominantly tied to APOGEE). 
A small offset of $0.05$~dex in metallicity and alpha abundance is measured, which should be taken into account when comparing \minesweepercat{} parameters to literature parameters.

\begin{figure*}
    \centering
    \includegraphics[width=1.8\columnwidth]{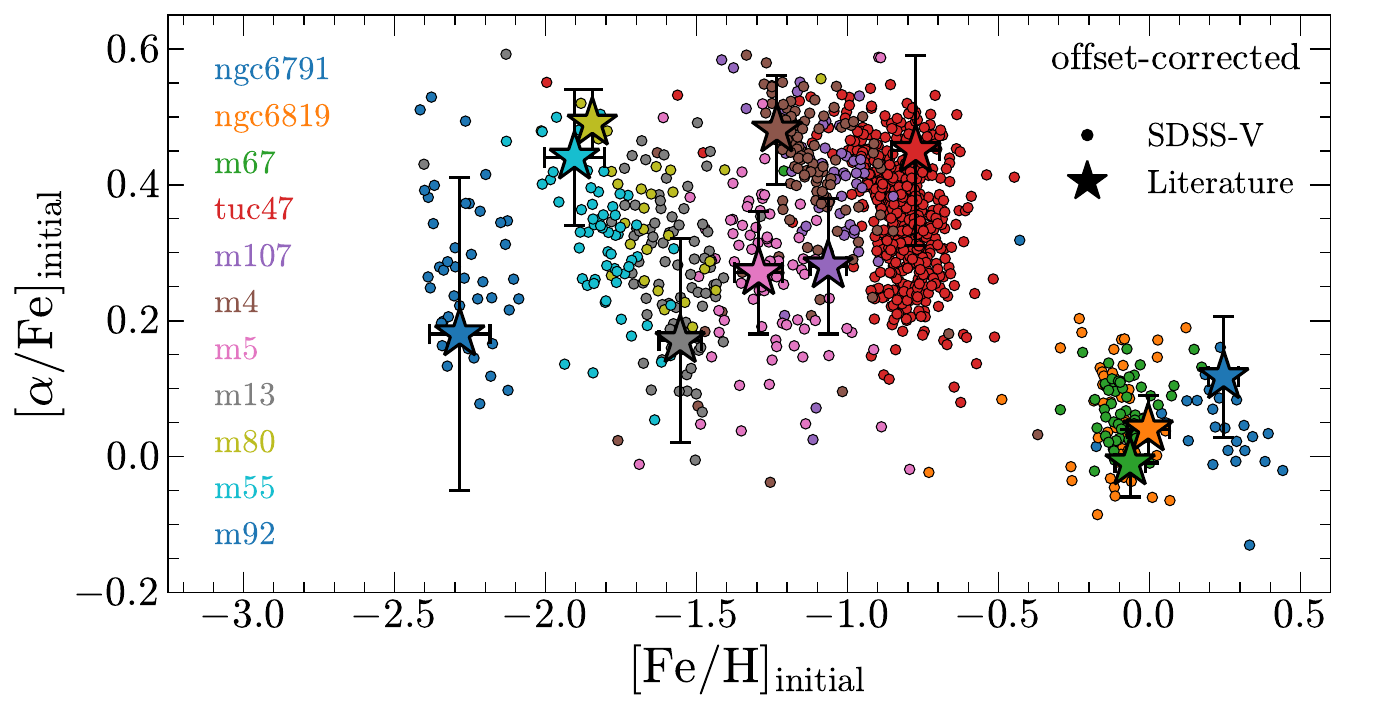}
    \caption{Metallicity versus alpha abundance for stars in our catalog belonging to calibration star clusters. 
    Mean values from the literature for each cluster are shown as stars, with corresponding $1\sigma$ error bars. 
    A constant offset in metallicity and alpha abundance has been subtracted off to bring the measurements onto the same scale (see Figure~\ref{fig:cluster_means}).}
    \label{fig:tw_cluster}
\end{figure*}

Figure~\ref{fig:tw_cluster} shows the metallicity-alpha plane for individual cluster stars observed by SDSS-V. 
Note that although error bars are not shown, these stars have typical statistical uncertainties of $\approx 0.1$~dex in metallicity and alpha abundance. 
Star symbols indicate the mean metallicity and alpha (typically magnesium) abundance for each cluster, along with the reported uncertainties from the literature. 
A constant offset on each axis (shown in Figure~\ref{fig:cluster_means}) has been removed to bring the \minesweepercat{} and literature values onto the same zeropoint scale. 
\minesweepercat{} [Fe/H] are tightly grouped around the literature values for each cluster, as expected for chemically homogeneous populations. 
% The [$\alpha$/Fe] matches for the metal-rich open clusters with [Fe/H]$\gtrsim-0.5$, but the scatter is larger for the more metal-poor globular clusters. This is expected because multiple stellar populations cause intrinsic Mg scatter.
This figure demonstrates that \minesweepercat{} recovers not only the mean cluster parameters, but also produces reliable metallicities and alpha abundances for individual stars. 

A key strength of cluster calibration is that we can evaluate how stars with identical metallicities are fit over a range of stellar parameters and evolutionary phases. 
Spectroscopic abundances can be susceptible to biases as a function of stellar parameters, often requiring re-calibrations at the catalog level \cite[e.g.,][]{Meszaros2025}. 
When studying halo structures that are spread over a wide range of distances (e.g., the Sagittarius Stream, $\S$\ref{sec:sgr}), these biases complicate interpretations by introducing artificial variations and gradients.
In principle, the use of photometry, astrometry, and isochrone priors in \minesweeper{} should greatly alleviate this problem, since the stellar parameters are independently constrained. 
Over a wide range of metallicities, there are no strong trends in the \minesweepercat{} metallicities as a function of evolutionary phase. 
There is a slight increase in metallicity when approaching the turnoff above $\log{g} \gtrsim 4$: this may be due an insufficient treatment of diffusion in the isochrone models. 
Although the effect is small, if precision abundances are required it is recommended to limit the analysis to red giant stars with $\log{g} \lesssim 3.5$. 

% \begin{figure}
%     \centering
%     \includegraphics[width=\columnwidth]{trend_logg_feh.pdf}
%     \caption{\minesweeper{} metallicity versus surface gravity for stars in calibration clusters for which we have more than 50 members. We use the initial isochrone metallicity returned by \minesweeper{}, which should be near-constant in a cluster regardless of evolutionary phase. 
%     }
%     \label{fig:cluster_logg}
% \end{figure}

\subsection{Cross-Comparison with Other Catalogs}

APOGEE \citep{Majewski2017} is a high-resolution near-infrared survey that delivers stellar parameters which have been extensively calibrated and validated \citep{GarciaPerez2016, Jonsson2018}. 
We cross-match our \minesweepercat{} catalog to the APOGEE DR17 \texttt{allStar} catalog \citep{GarciaPerez2016, Abdurro'uf2022}, finding $\gtrsim 1100$ stars in common. 

\begin{figure}
    \centering
    \includegraphics[width=\columnwidth]{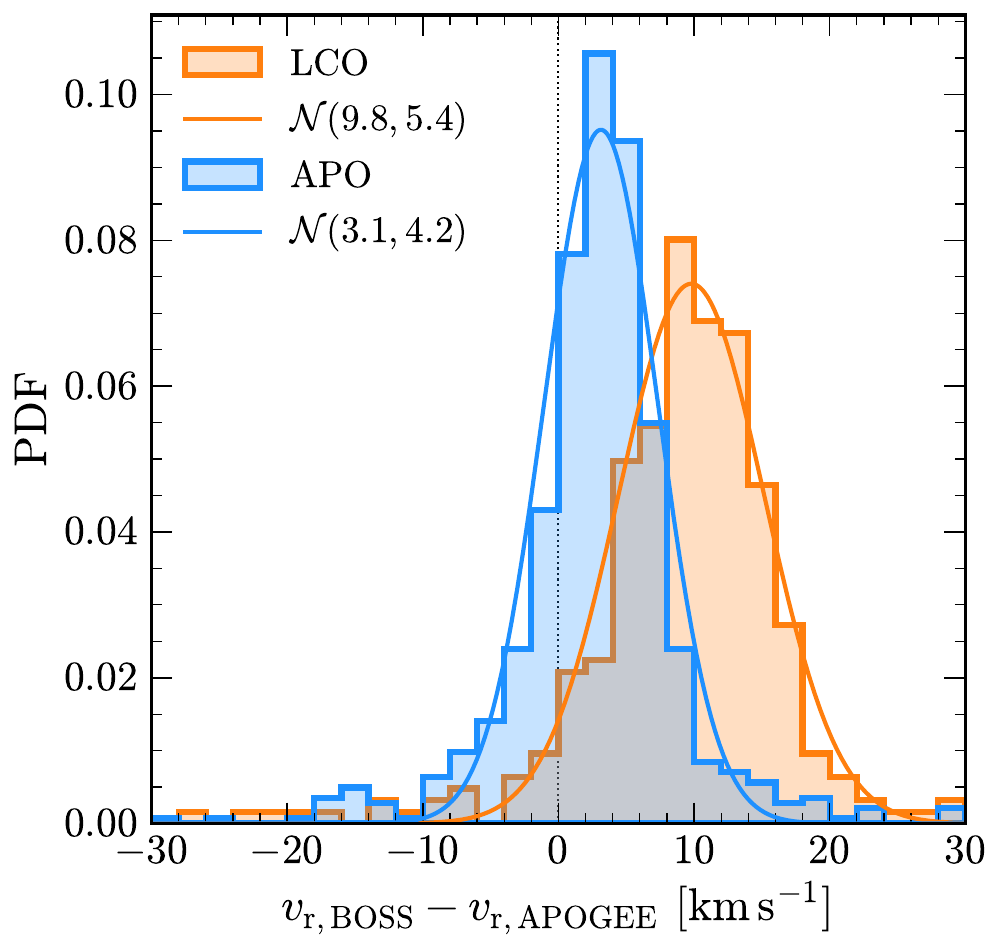}
    \includegraphics[width=\columnwidth]{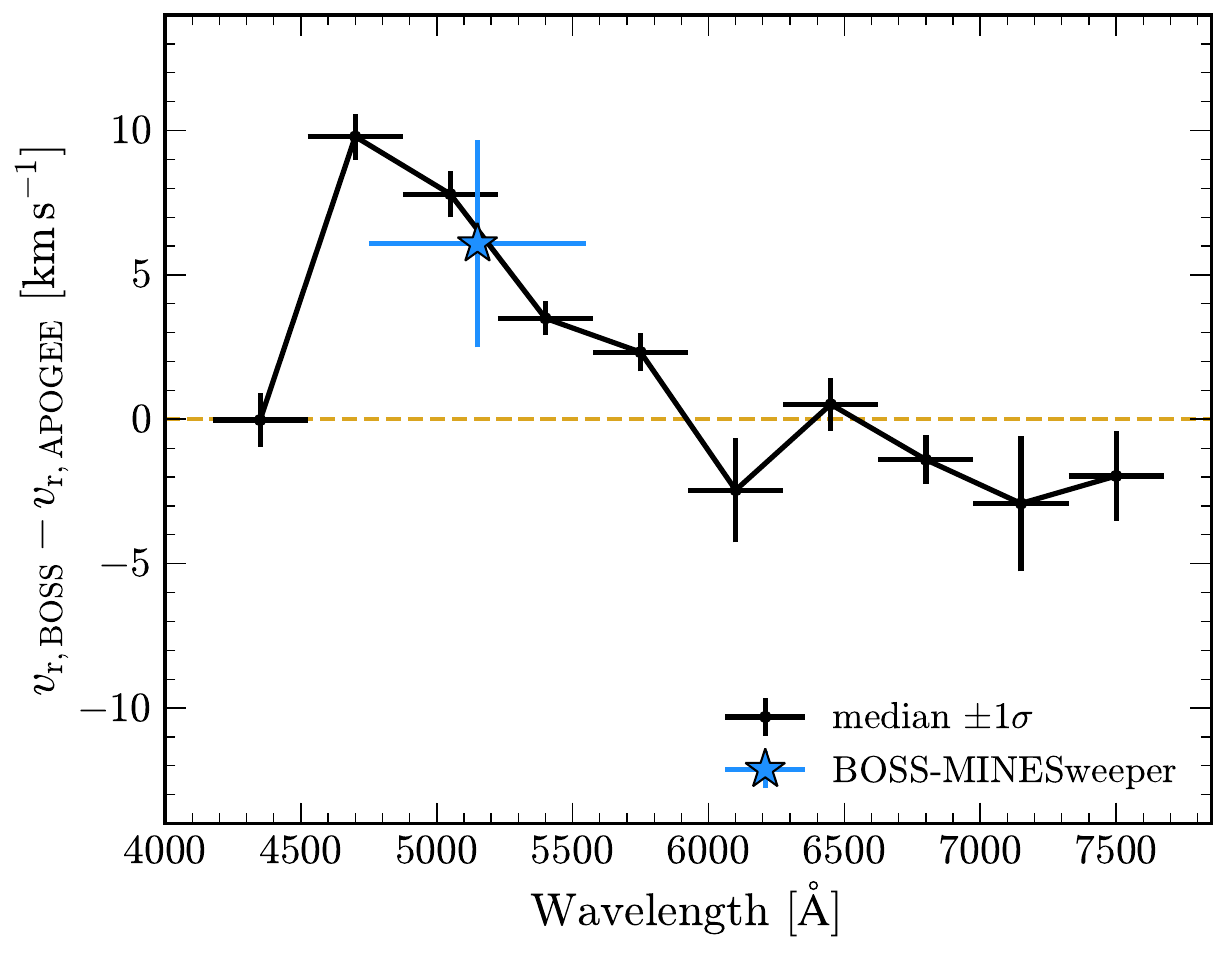}
    \caption{\textbf{Top:} Difference between \minesweepercat{} radial velocities and APOGEE radial velocities for stars observed by both surveys.
    The smooth curves show Gaussian distributions with the indicated mean and standard deviation.
    \textbf{Bottom:} RVs measured in narrow wavelength bins, relative to the APOGEE RV, for a subset of high-SNR BOSS spectra.
    The black curve is the median trend (including APO and LCO data), and the blue star shows the median \minesweepercat{} offset and wavelength regime. 
    }
    \label{fig:bossms-apogee-rv}
\end{figure}

Figure~\ref{fig:bossms-apogee-rv} shows the distribution of differences between \minesweepercat{} and APOGEE radial velocities. 
A small subset of stars have $\gtrsim 30$\,\kms{} discrepant RVs between the surveys and are not shown on this figure --- these are likely binary stars. 
For the remaining stars, there is an offset between \minesweepercat{} and APOGEE RVs, $\approx 3$~\kms{} for APO and $\approx 9$~\kms{} for LCO. 
This difference originates in the BOSS wavelength calibration in DR19, which is known to have small offsets at bluer wavelengths due to the relative paucity of arc and sky emission lines in that wavelength regime. 
This is illustrated in the bottom panel of Figure~\ref{fig:bossms-apogee-rv}, in which we fit the BOSS RV in narrow wavelength bins using the \texttt{doppler} code \citep{Ting2019, Nidever2021}, keeping the stellar parameters fixed to their \minesweepercat{} values. 
At wavelengths $\lesssim 5500$~\AA{}, the BOSS RVs are systematically redshifted compared to APOGEE, suggesting that the wavelength solution is biased in this regime. 
Since \minesweeper{} fits a narrow wavelength range $4750-5500$\,\AA{}, the wavelength solution is net redshifted for \minesweepercat{} spectra. 

The default SDSS-V RV product is fit using \texttt{pyXCSAO} \citep{Kurtz1992, Kounkel2022} over a much wider wavelength range, producing RVs that are more consistent with APOGEE.  
Since the public DR19 catalog only contains APO data, which have a relatively small RV offset, we publish the \minesweepercat{} RVs as-is. 
Users requiring more precise RVs should use the \texttt{XCSAO\_RV} column from the SDSS-V \texttt{spAll} table directly. 
For future data releases, the reduction pipeline is expected to undergo further improvements, particularly for LCO data. 
If these offsets persist, future iterations of the \minesweepercat{} catalog will report `re-centered' radial velocities that are post-processed to match the APOGEE scale. 

\begin{figure*}
    \centering
    \includegraphics[width=\textwidth]{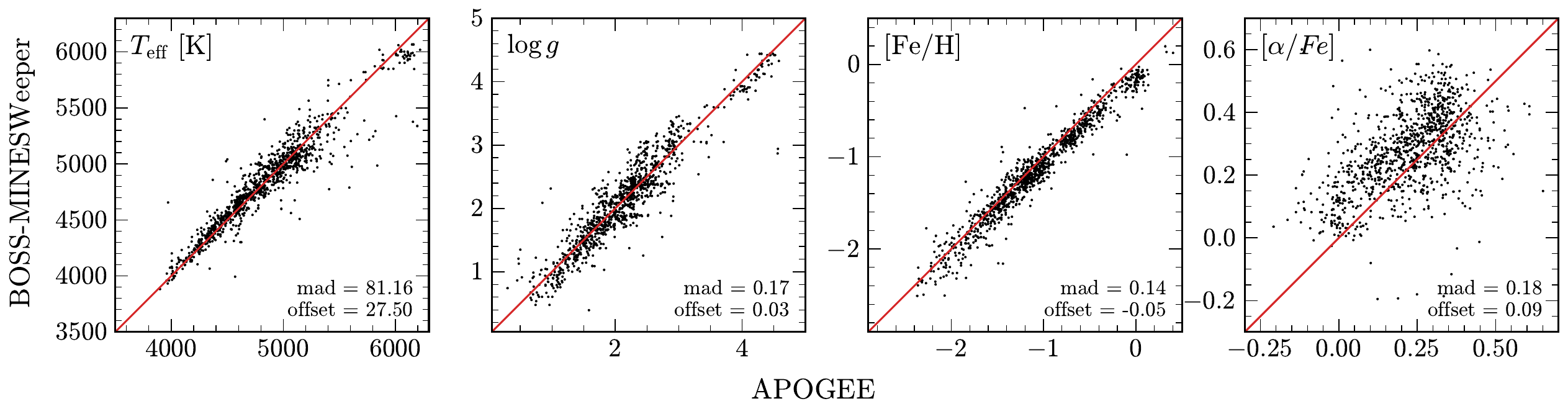}
    \caption{Cross-validating \minesweepercat{} stellar parameters using stars shared with the APOGEE DR17 catalog.
    The (centered) median absolute deviation and median bias are indicated.}
    \label{fig:bossms-apogee}
\end{figure*}

Figure~\ref{fig:bossms-apogee} shows parameter comparisons between \minesweepercat{} and APOGEE for temperature, surface gravity,  metallicity, and alpha abundance. 
As in the previous section, we use the APOGEE magnesium abundance as our reference for [$\alpha$/Fe], since the \minesweepercat{} wavelength range is most sensitive to Mg. 
There is overall excellent agreement between our \minesweepercat{} results and the APOGEE parameters, despite the BOSS spectra having substantially lower resolution ($R \approx 1800$ compared to APOGEE's $R \approx 22,000$) and lower typical SNR. 
\minesweeper{} slightly underestimates [$\alpha$/Fe] compared to the APOGEE scale, which should be taken into account if comparing parameters across surveys. 
Later in this work, Figure~\ref{fig:followup} extends the metallicity comparison to the extremely metal-poor regime, finding excellent agreement between \minesweepercat{} and follow-up high-resolution spectroscopy. 

\begin{figure}
    \centering
    \includegraphics[width=\columnwidth]{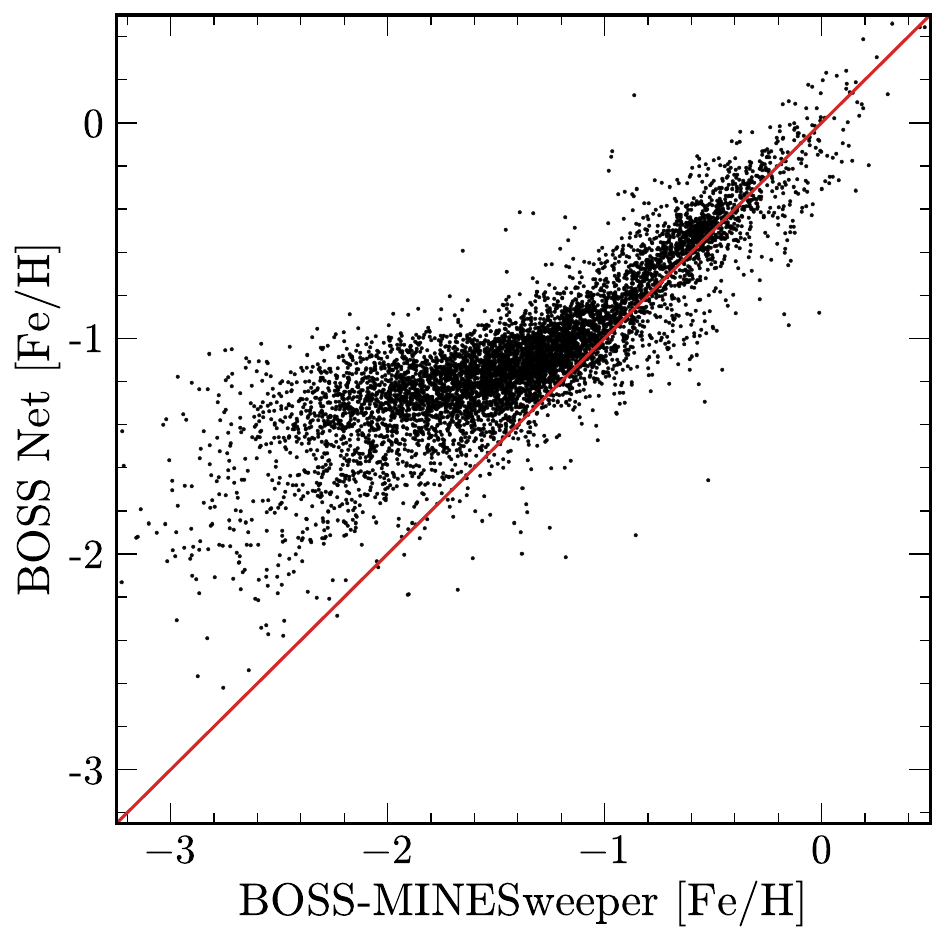}
    \caption{Comparison between metallicities measured by \minesweeper{} and those measured by the machine learning-based BOSS Net code that is run on all SDSS-V spectra.
    There is good agreement for most stars, but below [Fe/H]~$\lesssim -1.0$ BOSS Net systematically overestimates metallicities compared to \minesweeper{}.
    }
    \label{fig:bossms-bossnet}
\end{figure}

Figure~\ref{fig:bossms-bossnet} compares \minesweepercat{} metallicities to those estimated by the BOSS Net utility, which is the default parameter pipeline for SDSS-V \citep{Sizemore2024}. 
BOSS Net is a neural network that performs `label transfer': trained on a large sample of BOSS spectra with known parameters, it estimates stellar parameters from the BOSS spectrum directly. 
The advantage of this methodology is that BOSS Net is extremely fast to evaluate, and is consequently run on every star observed by SDSS-V. 
As shown in Figure~\ref{fig:bossms-bossnet}, BOSS Net and \minesweeper{} return comparable metallicities above [Fe/H]~$\gtrsim -1.0$. 
However, at metallicities lower than this, the BOSS Net metallicities bottom out, whereas \minesweeper{} continues to deliver reliable metallicities (see the earlier comparisons to star clusters and APOGEE). 

This test highlights a pitfall of ML-based predictive models, which is that they tend to regress towards the mean, particularly when applied to stars in the extremes of a parameter distribution \citep[see][for a pedagogical example]{Hogg2024}. 
Tools like BOSS Net are useful and accurate over the majority of the parameter space, and their speed enables their application to millions of stars across SDSS-V. 
However, for the subset of distant and metal-poor stars that are most interesting to the halo survey, a specialized tool like \minesweeper{} is required. 

\section{The \minesweepercat{} Science Showcase}\label{sec:science}

Several projects are underway to utilize the wealth of data delivered by SDSS-V.
Although the primary purpose of this paper is to present and describe the \minesweepercat{} catalog, in this section we summarize some science highlights from these ongoing projects. 

\subsection{Overview and Targeting Efficiency}\label{sec:mscat}

\begin{figure}
    \centering
    \includegraphics[width=\columnwidth]{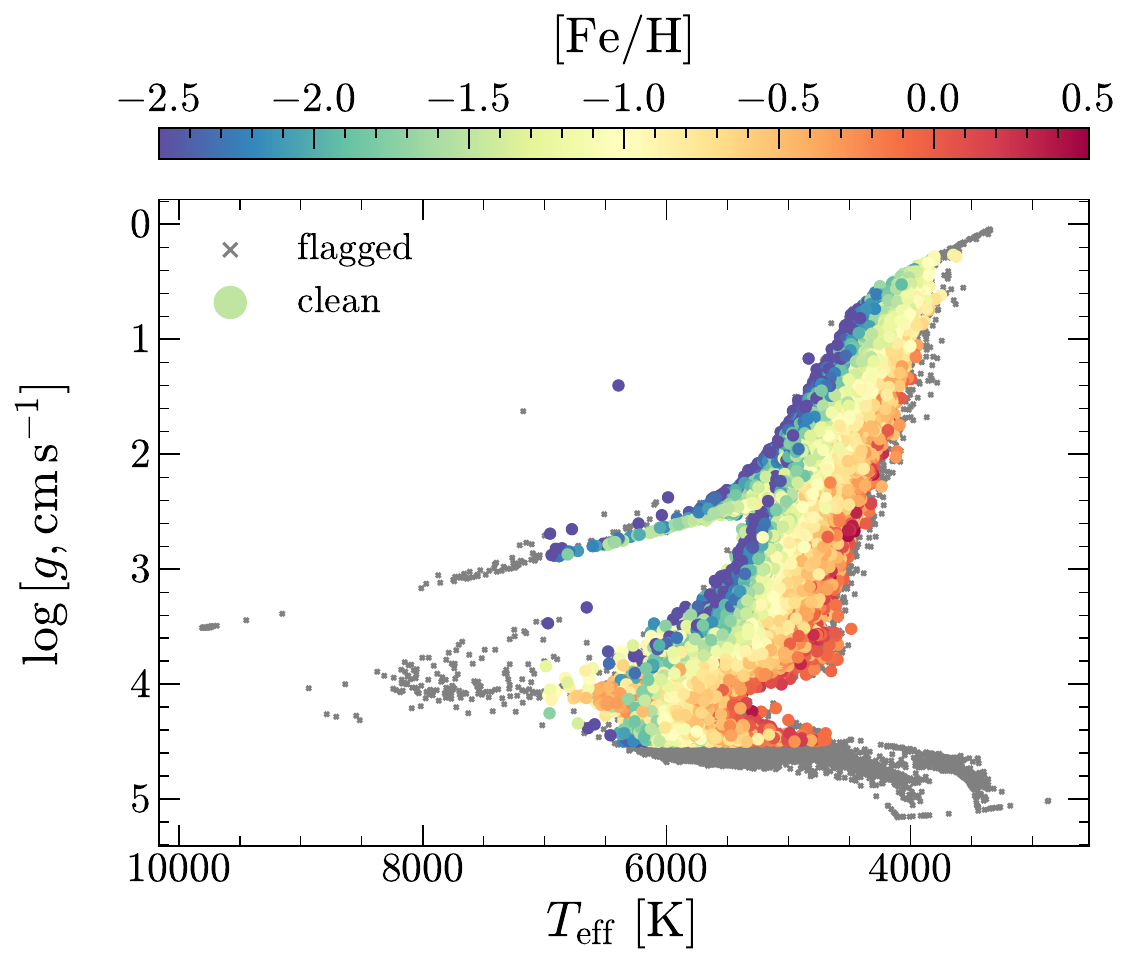}
    \includegraphics[width=\columnwidth]{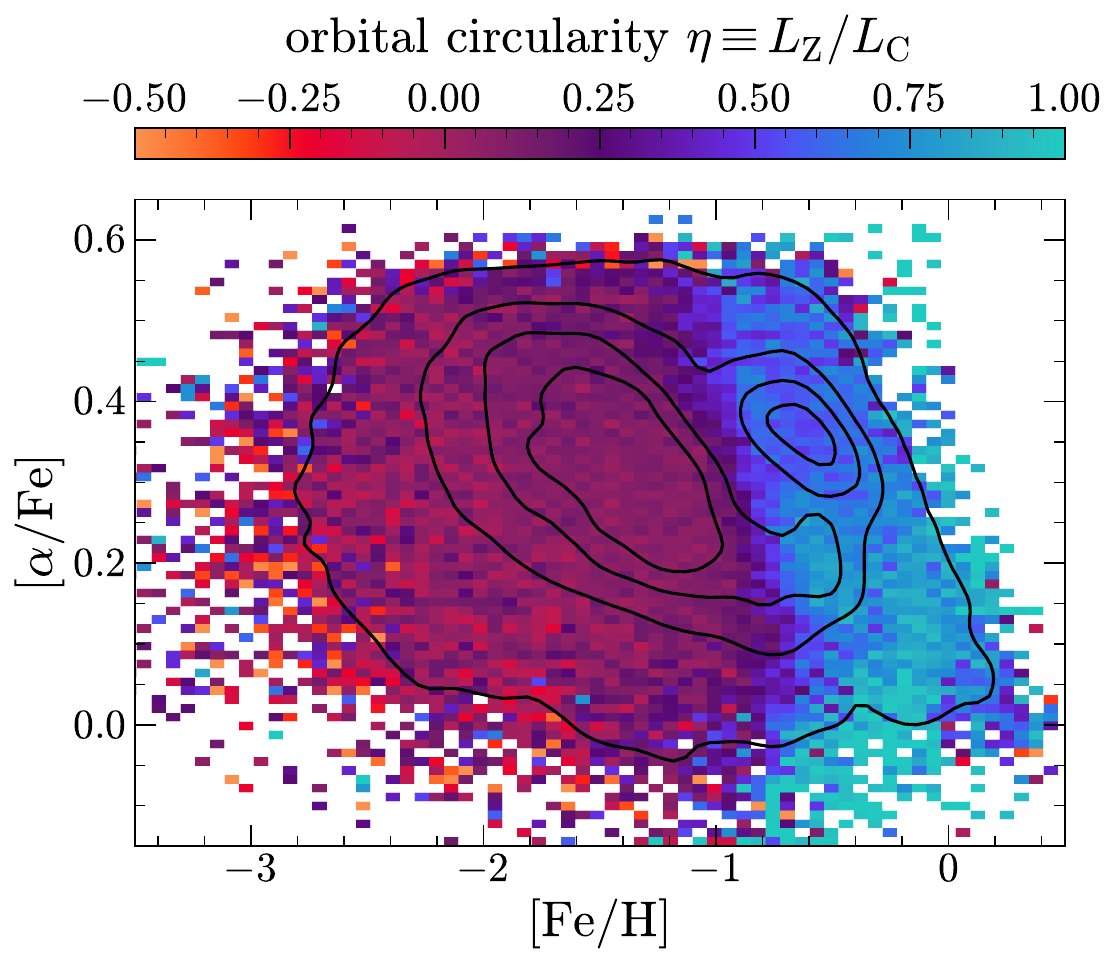}
    \caption{\textbf{Top:} Kiel diagram of stars in the catalog, colored by metallicity. 
    The influence of the isochrone constraint in \minesweeper{} is evident here.  
    Although most stars are old --- belonging to the thick disk and halo --- a small population of younger metal-rich stars from the thin disk are visible near the main sequence turnoff. 
    Grey crosses denote stars in regions of the parameter space for which \minesweeper{} parameters are deemed unreliable. 
    \textbf{Bottom:} Metallicity-$\alpha$ distribution of stars in the catalog, colored by the mean orbital circularity. 
    Contours denote the density of stars in this space.}
    \label{fig:kiel}
\end{figure}

Key \minesweepercat{} parameters are summarized in Figure~\ref{fig:kiel}. 
The Kiel diagram --- the spectroscopic Hertzsprung–Russell (HR) diagram --- illustrates a key strength of the \minesweeper{} code, namely that solutions are constrained to lie on physically-plausibly regimes of the HR diagram. 
Grey points illustrate stars that are flagged as having unreliable \minesweepercat{} parameters. 
These include cool M and K dwarfs, and stars hotter than $T_\mathrm{eff} > 7000$~K, which have unreliable metallicities. 
The bottom panel shows the Tinsley-Wallerstein diagram --- the distribution of metallicities [Fe/H] and alpha abundances [$\alpha$/Fe] --- colored by orbital circularity (the ratio between the star's specific angular momentum $L_\mathrm{Z}$ and the angular momentum for a perfectly circular orbit with the same total energy $L_\mathrm{C}$; e.g., \citealt{Chandra2024}).
At solar metallicity, highly circular stars ($\eta \gtrsim 0.9$) in the cool low-$\alpha$ disk of the MW are visible. 
At lower metallicities, stars in the high-$\alpha$ disk with slightly more disordered orbits are visible  ($\eta \gtrsim 0.5$). 
Finally, the majority of the sample is composed of bona-fide halo stars at lower metallicities with markedly scattered orbits ($\eta \sim 0$ with large variance). 
% Most of these stars were likely deposited by the Gaia-Sausage Enceladus (GSE) merger, which is thought to have built up the majority of the MW's stellar halo \citep[e.g.,][]{Helmi2018, Belokurov2018a, Naidu2020, Naidu2021, Chandra2023a}. 

\begin{figure}
    \centering
    \includegraphics[width=\columnwidth]{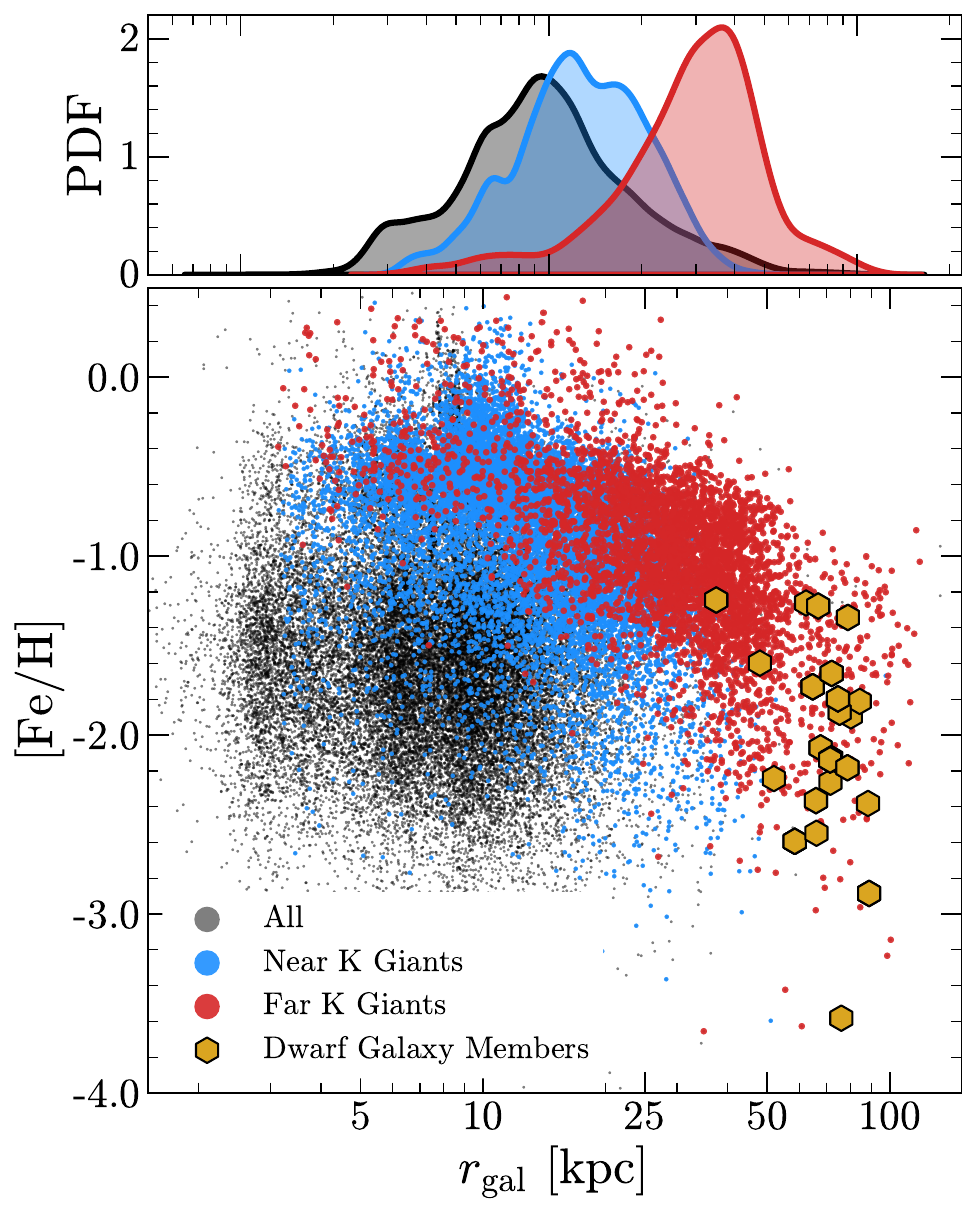}
    \caption{Performance of the distant halo selection carton. 
    Blue (red) points show giants targeted to lie beyond $d_\mathrm{helio} > 10~(30)$~kpc. 
    The top panel shows the marginal distribution of $r_\mathrm{gal}$. 
    Gold points show a cross-match to the dwarf spheroidal galaxy member catalog from \cite{Walker2023}. 
    Our parallax and color-based target selection is highly efficient at identifying the most distant stars in the MW halo.}
    \label{fig:kg}
\end{figure}

Figure~\ref{fig:kg} showcases the efficiency of the target selection for distant giants. 
The `Near K Giant' carton targets stars beyond $\gtrsim 10$~kpc, and the high-priority `Far K Giant' targets stars beyond $\gtrsim 30$~kpc. 
Both of these selections are extremely efficient, with minimal contamination from nearby dwarf stars. 
There is a marked trend with distance, caused by the IR color-based selection: for a given selection, the more distant stars are more metal-poor. 
Efforts to characterize the 3D metallicity distribution of the halo will need to correct for this selection effect, which can be done by forward-modelling the IR color cuts. 
Figure~\ref{fig:kg} also shows stars cross-matched to the dwarf spheroidal galaxy catalog of \cite{Walker2023}.
The \minesweepercat{} catalog incidentally contains $> 5$ members in Bo{\"o}tes~I, Carina, Sextans, and Ursa~Minor~I. 
These stars are among the most distant and metal-poor in the catalog, highlighting the ability of the \minesweepercat{} distances and metallicities to identify new members of known dwarfs, or even discover new dwarfs entirely. 

\begin{figure}
    \centering
    \includegraphics[width=\columnwidth]{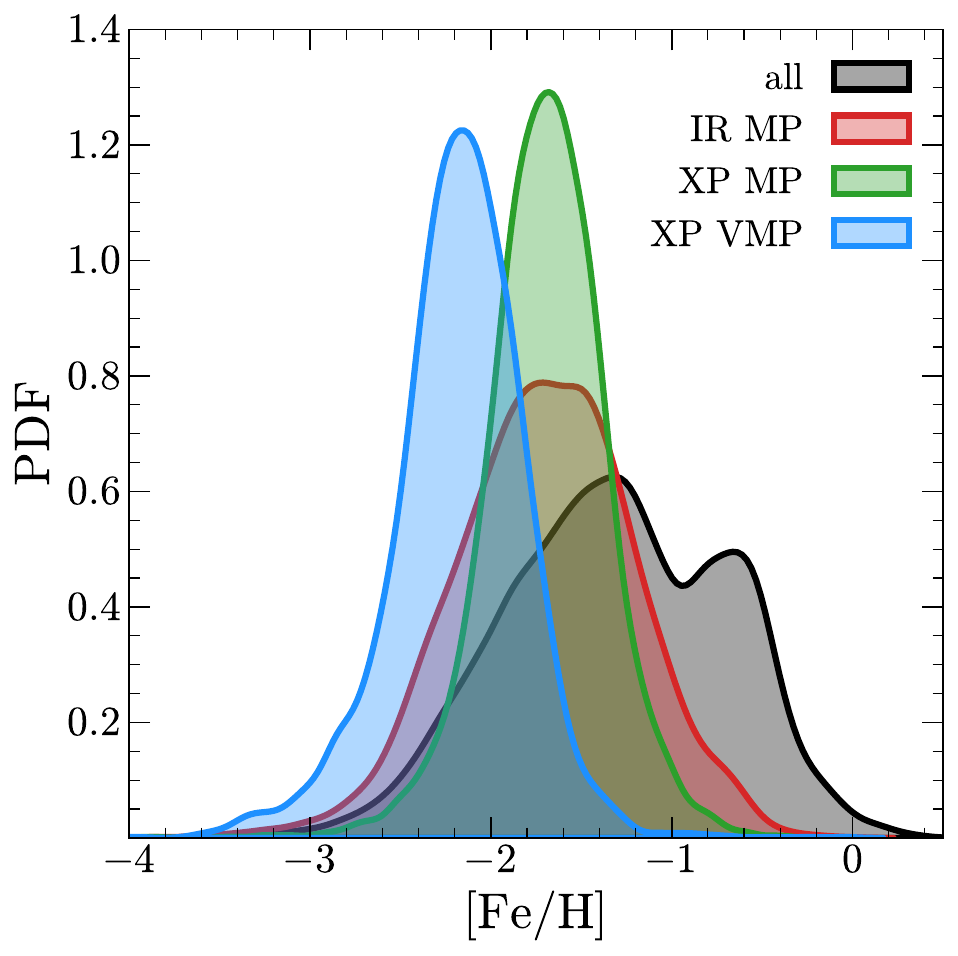}
    \caption{Performance of the metal-poor target selection, which uses \textit{Gaia} XP-based metallicities to identify metal-poor ([Fe/H]~$< -1.5$, green) and very metal-poor stars ([Fe/H]~$< -2.0$, blue).
    The red distribution shows stars from an older selection that relies on infrared colors. 
    }
    \label{fig:mp}
\end{figure}

Figure~\ref{fig:mp} evaluates several cartons designed to target metal-poor stars. 
The high-priority XP-VMP and XP-MP cartons target stars with [Fe/H]$< -2.0$ and [Fe/H]$< -1.5$ respectively, based on low-resolution prism spectra from \textit{Gaia} DR3 \citep{Andrae2023}. 
These cartons have turned out to be very efficient at identifying metal-poor stars: so far $\approx 2500$ stars targeted in the VMP carton have [Fe/H]$< -2.0$, and $\gtrsim 9500$ stars targeted in the VMP and MP cartons have [Fe/H]$< -1.5$, with an overall efficiency of $\approx 70\%$ in targeting VMP stars.
Including the XP information is about 10${\times}$ more efficient at identifying stars with [Fe/H] $< -2$ than a random selection of halo stars.
The IR-MP selection using 2MASS and WISE photometry has similar efficiency to the XP-MP carton, and works better than the XP selection in more reddened regions close to the MW disk. 

\begin{figure*}
    \centering
    \includegraphics[width=\textwidth]{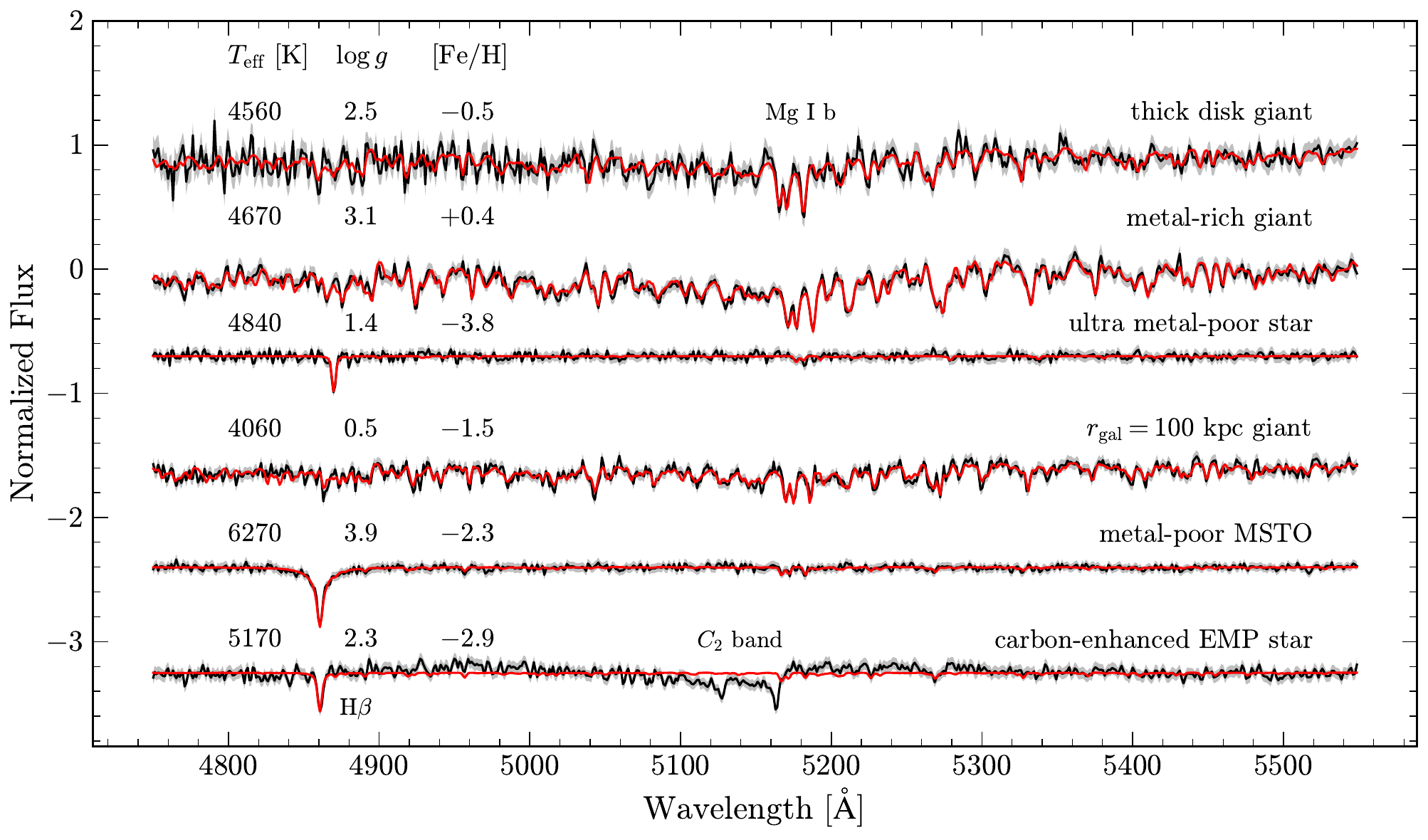}
    \caption{Showcase of various interesting stars identified from the \minesweeper{} catalog. The BOSS spectra are shown in black, and the model fit is shown in red. \minesweeper{} stellar parameters are indicated for each star.
    The third star was discovered in the \minesweepercat{} catalog and is one of the most metal-poor stars ever found (Ji et al, in prep).
    The last star shows a strong $C_2$ feature that is not captured by our spectral models, revealing it to be a carbon--enhanced extremely metal-poor star. 
    The \ion{Mg}{1}\,b triplet and H$\beta$ line are also indicated.}
    \label{fig:showcase} 
\end{figure*}

Figure~\ref{fig:showcase} illustrates several stars selected from the \minesweepercat{} catalog that demonstrate the diversity of spectra in the SDSS-V halo survey. 
Within the disk of the MW, we can identify typical thick disk giants, as well as extremely metal-rich giants residing in the inner galaxy \citep[e.g.,][]{Rix2024, Horta2025}. 
\minesweeper{} also excels at breaking the degeneracy between hot stars and cooler metal-poor stars, which have similar spectra but markedly different broadband spectral energy distributions. 
The ultra metal-poor star shown in Figure~\ref{fig:showcase} --- an SDSS-V discovery --- has turned out to be one of the most metal-poor stars ever found (Ji et al, in prep).
The figure also shows an example of a metal-poor main sequence turn-off (MSTO) star, which can be used to derive precise isochrone-based ages, timing the assembly history of the early galaxy \citep[e.g.,][]{Xiang2022, Woody2025}. 

\subsection{High-Resolution Spectroscopic Followup}

% \textbf{Alex to add text and references.}

One of the primary goals of the \minesweepercat{} effort is to identify interesting metal-poor stars that warrant follow-up observations with high-resolution spectrographs.
These chemically-primitive stars hold the key to the earliest nucleosynthetic events in our Galaxy \citep[e.g.,][]{Beers2005, Frebel2015, Bonifacio2025}.
The \minesweepercat{} parameters are essential to reliably identify these candidates, as the short 15-minute visits in SDSS-V result in lower signal-to-noise spectra that need the photometric information to obtain reliable metallicities. 
\minesweepercat{} alpha abundances can be used to identify low-alpha metal-poor (LAMP) stars, which are expected to have been born in ancient low-mass dwarf galaxies consumed by the Milky Way \citep[e.g.,][]{Robertson2005,Cunningham2022}. 

Within the SDSS-V collaboration, an extensive high-resolution follow-up program is underway. 
These observations are primarily conducted with the MIKE spectrograph on the 6.5m Magellan Clay telescope in Chile \citep{Bernstein2003}, as well as the ARCES spectrograph on the Astrophysical Research Consortium 3.5m telescope at APO \citep{Wang2003}. 
Targets are selected to be extremely metal-poor ([Fe/H]~$\ll -3.0$) and/or low-alpha ([$\alpha$/Fe]~$\lesssim 0.1$). 

\begin{figure}
    \centering
    \includegraphics[width=\linewidth]{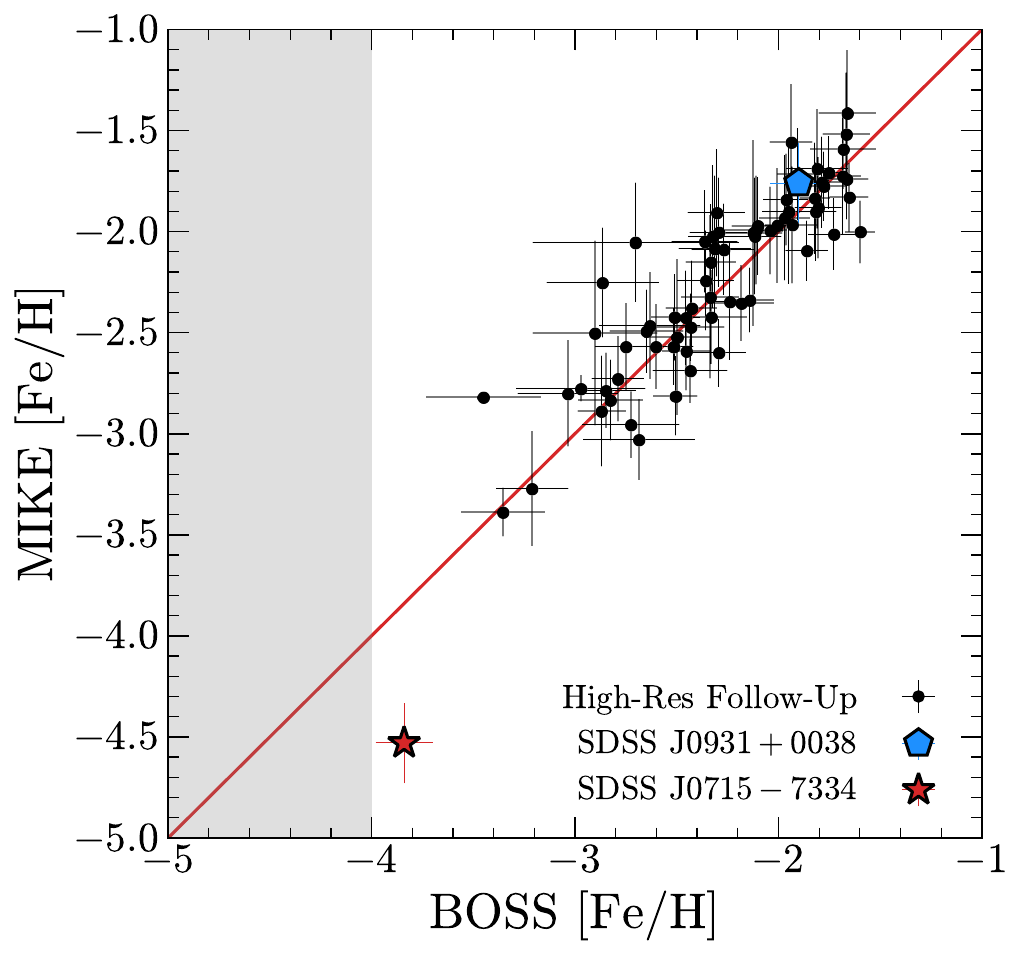}
    \caption{Comparison between BOSS \minesweeper{} metallicities and those derived from high-resolution MIKE spectra, for a subset of metal-poor stars. 
    \minesweeper{} metallicities are reliable down to the ultra metal-poor regime at [Fe/H]~$-4.0$, the lower limit of our model grids.
    }
    \label{fig:followup}
\end{figure}

Figure~\ref{fig:followup} shows 69 stars for which high-resolution spectra were obtained with the MIKE spectrograph (PIs: Alexander Ji and Danielle de Brito Silva). 
Metallicities were determined from the $R \approx 30,000$ MIKE spectra using the \texttt{LESSPayne} utility \citep{Ji2025lp}. 
There is overall excellent agreement between the BOSS and MIKE metallicities, all the way down to the lowest edge of the \minesweeper{} grid at [Fe/H]~$= -4.0$. 
BOSS has already discovered several ultra metal-poor (UMP) stars that have been confirmed with high-resolution spectroscopy to have [Fe/H]~$< -4.0$. 

In Figure~\ref{fig:followup} we highlight two stars that turned out to be particularly interesting. 
SDSS~J$09341+0038$ is a moderately metal-poor, low-alpha star which was discovered in the SDSS-V \minesweepercat{} data, and subsequently followed up and described in \cite{Ji2024}. 
This star exhibits a highly unusual nucleosynthetic signature, with some of the lowest ever measured abundances of elements like sodium, aluminum, potassium, and barium but unusually high abundances of iron-peak and trans-iron elements like zinc and strontium. 
The physical implication is that the metals in this star likely originated from a single nucleosynthetic source, probably a rotationally-powered supernova of an exceptionally massive star. 
This star highlights the limitations of current nucleosynthetic models in this mass regime, and more examples will probably be found in data from SDSS-V and other large surveys. 

A second discovery from the SDSS-V \minesweepercat{} catalog is SDSS~J$0715 - 7334$, described in Ji et al (in prep). 
This star was flagged as being close to the low edge of the \minesweeper{} metallicity grid at [Fe/H]~$= -4.0$, and followed up with MIKE. 
The high-resolution spectrum confirmed a UMP metallicity of [Fe/H]~$= -4.5$. 
However, unlike most UMP stars \citep{Frebel2015}, SDSS~J$0715 - 7334$ is also highly deficient in carbon and other metals. 
Overall, SDSS~J$0715 - 7334$ is likely the most metal-poor star --- in terms of total metal content --- discovered to date. 
This star is especially interesting due to its kinematics, which strongly suggest that the star was born in the LMC and subsequently captured by the Milky Way. 
This adds to a growing picture that dwarf galaxies like the LMC are deficient in carbon-enhanced metal-poor stars, suggesting that their nucleosynthetic history is quite different from that of larger galaxies like the MW \citep[e.g.,][]{Chiti2024}.

Several other interesting stars have been identified in the \minesweepercat{} catalog, and will be the subject of future work. 
As these catalogs become public in DR19 and future data releases, the broader community will hopefully continue to follow up chemically interesting stars with high-resolution spectroscopy. Low-metallicity stars in SDSS-V clearly host plenty of surprises. 

\subsection{Search for Substructures}

The stellar halo of the MW is filled with coherent substructures, from bound dwarf galaxies to stellar streams \citep[e.g.,][]{Irwin1990sxt, Grillmair2006, Helmi1999b, Helmi2008, Belokurov2006b, Belokurov2007a, Belokurov2007, Helmi2020, Naidu2020, Ibata2021, Malhan2022a, Bonaca2025}. 
These structures are typically identified as overdensities of stars in photometric surveys, with color-magnitude filtering being a useful tool to increase the contrast of distant structures. 
In the era of \textit{Gaia} parallaxes and proper motions, the \textit{kinematic} search for substructure has become a powerful tool to identify systems that have surface brightnesses too low to be detected via photometry \citep[e.g.,][]{Malhan2018, Torrealba2019}. 
These techniques become even more powerful with spectroscopy, since radial velocities and metallicities further increase the contrast of substructures against other halo stars. 
Spectroscopically identifying even a few bright, co-moving members of a dwarf or stream enables the proper motion-based filtering of the broader \textit{Gaia} catalog, revealing the entirety of the dwarf galaxy or stream \citep[e.g.,][]{Chandra2022, Aganze2025}. 

Figure~\ref{fig:kg} showed the metallicity-distance distribution of the \minesweepercat{} catalog, with members of known dwarf satellite galaxies from \cite{Walker2023} highlighted. 
These stars were not specially targeted, but were incidentally selected for observation in SDSS-V. 
The dwarf galaxy members stand out as among the most distant and metal-poor in the entire \minesweepercat{} catalog. 
These stars also have \minesweeper{} distances that are consistent with literature distances for the respective dwarfs. 
A blind search among $r_\mathrm{gal} > 30$~kpc and [Fe/H]$< -2.0$ stars independently reveals several members in the Bo{\"o}tes~I and Sextans dwarf galaxies. 
The \minesweepercat{} catalog will be particularly useful to find members at large separations from known dwarf galaxies, where coverage from targeted spectroscopic follow-up can be sparse or absent. 

Due to their large spatial extent, stellar streams are the most common substructure that is identified by kinematic search techniques \citep[e.g.,][]{Chandra2022, Aganze2025}. 
Cross-matching the \minesweepercat{} catalog to the compilation of stream stars from \cite{Bonaca2025}, we find $\approx 1300$ members of the Sgr stream, and $\approx 200$ members of other known streams. 
The following streams have more than five stars observed in SDSS-V: 300S, ATLAS-Aliqa Uma, GD-1, Indus, Kwando, New-25, and Orphan-Chenab \citep{Grillmair2006, Belokurov2007a, Shipp2018, Fu2018, Li2021aau, Ibata2024}. 
Many more can be expected as the survey fills in regions of the sky (see Figure~\ref{fig:targets-sky}). 
These stars can be used to map the metallicity and alpha abundance structure of streams, and also to select high-confidence members for spectroscopic follow-up observations with high-resolution instruments \citep[e.g.,][]{Li2022b, Ji2022, Usman2024}. 
Future work will also perform a blind search for kinematic substructures in the SDSS-V data, potentially uncovering new stellar streams in the outer halo. 

\subsection{The Sagittarius Stream}\label{sec:sgr}

The Sagittarius Stream (Sgr) is by far the most massive stellar stream around the Milky Way, and spans hundreds of degrees across the sky \citep[e.g.,][]{Majewski2003}.
Sgr originates from the partially-disrupted Sagittarius dwarf galaxy, and this progenitor is thought to have originally had a stellar mass $\sim 10^8\,M_\odot$, and a dark matter halo mass $\gtrsim 10^{10}\,M_\odot$ \citep{Penarrubia2010, Gibbons2017, Laporte2018, Vasiliev2021a}. 
Although Sgr stars are not explicitly targeted in SDSS-V, the stream dominates the halo beyond $r_\mathrm{gal} \gtrsim 20$~kpc, and thousands of Sgr stars are consequently observed in our distant halo cartons. 
SDSS-V provides a unique all-sky perspective of the Sgr stream, and our catalog represents the largest all-sky sample of Sgr members with spectroscopic metallicity and alpha abundances. 

\begin{figure}
    \centering
    \includegraphics[width=\linewidth]{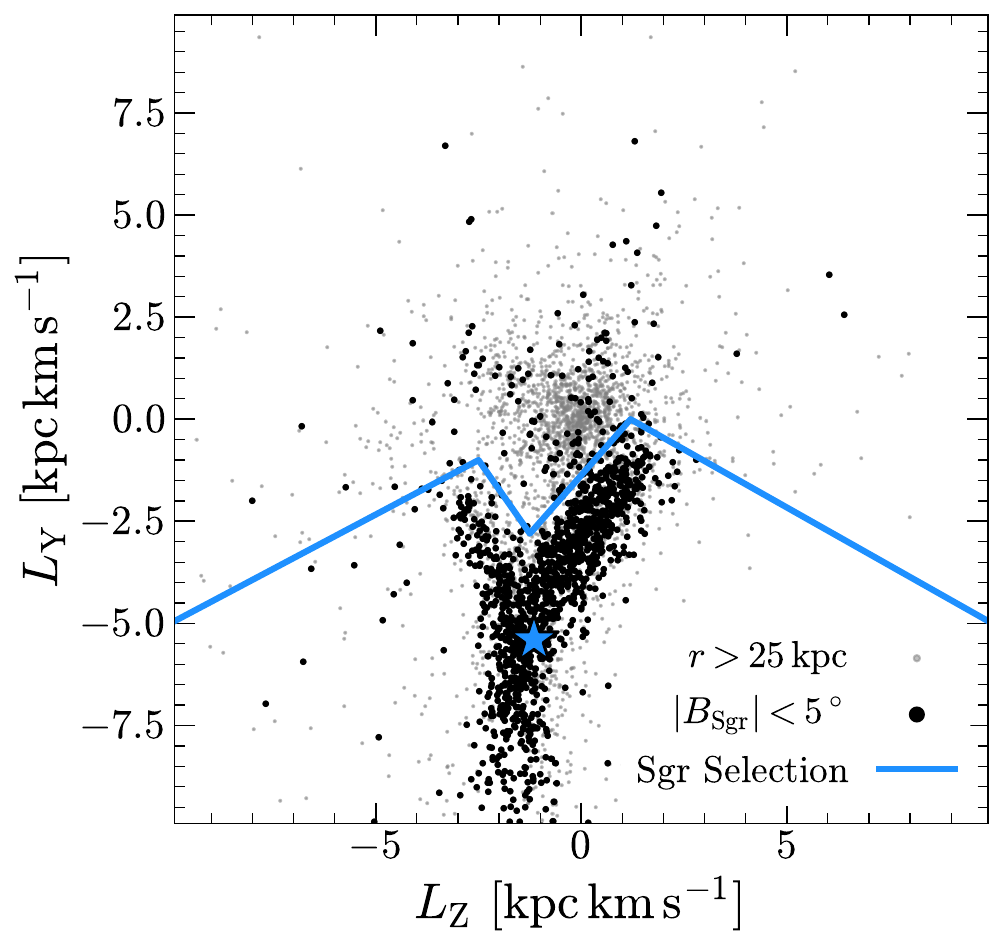}
    \includegraphics[width=\linewidth]{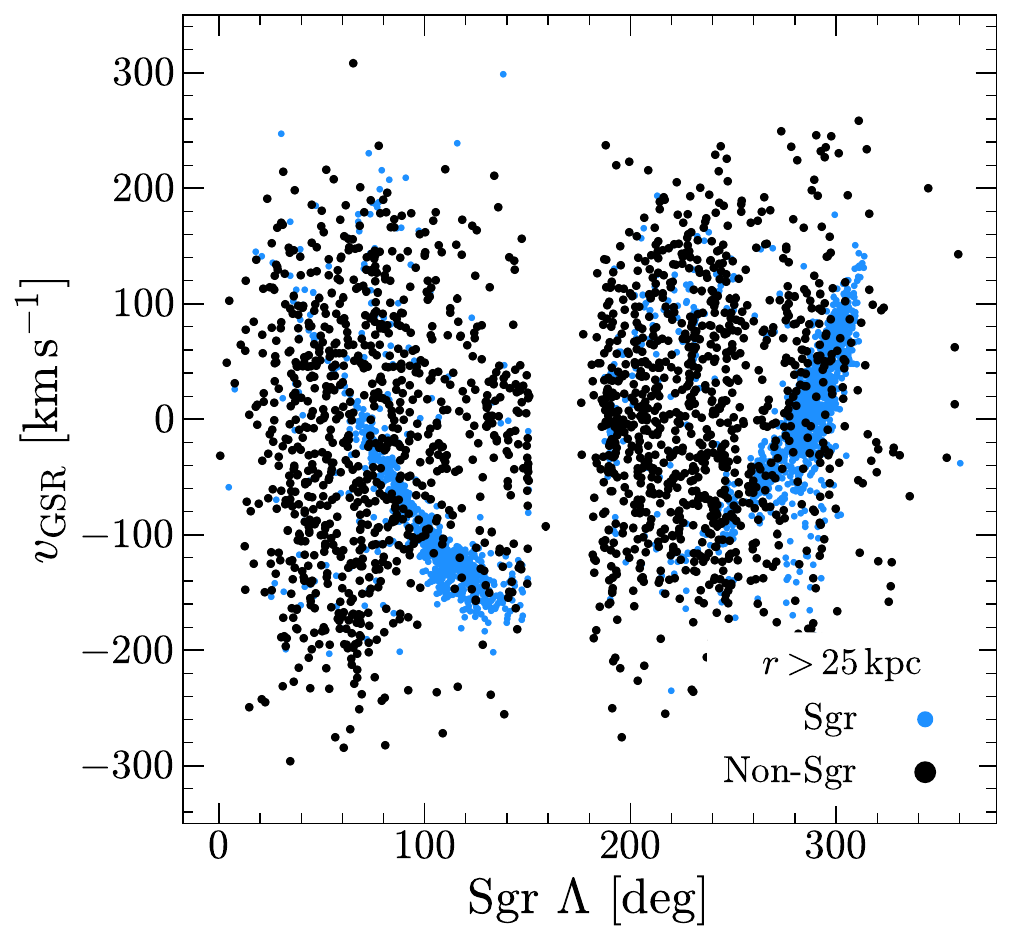}
    \caption{The Sagittarius Stream (Sgr) in SDSS-V. 
    \textbf{Top:} Selection of Sgr members in angular momentum space. Stars lying very close to the orbital plane of Sgr are used to define a polygon that captures most likely Sgr members. 
    \textbf{Bottom:} The Galactocentric radial velocity of stars beyond 25~kpc as a function of longitude along the Sgr stream, with angular momentum-selected Sgr stars highlighted in blue.
    Our selection is effective at excising Sgr stars from the field halo. }
    \label{fig:sgr}
\end{figure}

Due to the unique polar orbit of the Sgr stream, its members occupy a unique space in the angular momentum distribution of halo stars. 
\cite{Johnson2020a} present a straightforward kinematic selection to isolate Sgr members, based on data from the H3 Survey: $L_\mathrm{Y} < 2.5 - 0.3 \times L_\mathrm{Z}$, where the angular momenta are in units of \lunit. 
The top panel of Figure~\ref{fig:sgr} shows this angular momentum space for distant stars in our catalog, with stars close to the Sgr orbital track highlighted. 
Although the linear cut suggested by \cite{Johnson2020a} is sufficient for stars within the H3 footprint, a more complex cut is required to isolate Sgr members across the sky (blue line in Figure~\ref{fig:sgr}.
We use stars lying close to the Sgr orbital plane to define a polygon selection passing through the following $L_\mathrm{Z}-L_\mathrm{Y}$ points: $[(-10, -5), (-2.5, -1), (-1.25, -2.8), (1.2, 0), (10, -5)]$, where the units are \lunit. 
We mark a star as an Sgr candidate if it lies in that angular momentum box, is within $30^\circ$ of the Sgr orbital plane on-sky, and is beyond $r_\mathrm{gal} > 10$~kpc. 
$\approx 50\%$ of the stars in our catalog beyond $25$~kpc satisfy this cut, underscoring how dominant Sgr is at these distances. 
The angular momentum of the Sgr remnant is shown with a star in Figure~\ref{fig:sgr}. 
It is clear that the stream debris span a wide range of angular momenta, and consequently it is insufficient to select stream members using a spherical angular momentum cut centered around the remnant's angular momentum \citep[e.g.,][]{Penarrubia2021, Petersen2021, Limberg2023, Chandra2025}.  

The bottom panel of Figure~\ref{fig:sgr} shows the Galactocentric radial velocity of stars beyond $25$~kpc along the Sgr orbital longitude, colored by our Sgr selection. 
The main Sgr track forms a tight sequence on this plot. 
Furthermore, no significant overdensities are visible in the non-Sgr sample. 
Therefore, the Sgr flag included with our catalog can be used to isolate field halo stars for any scientific analyses. 

\begin{figure}
    \centering
    \includegraphics[width=\linewidth]{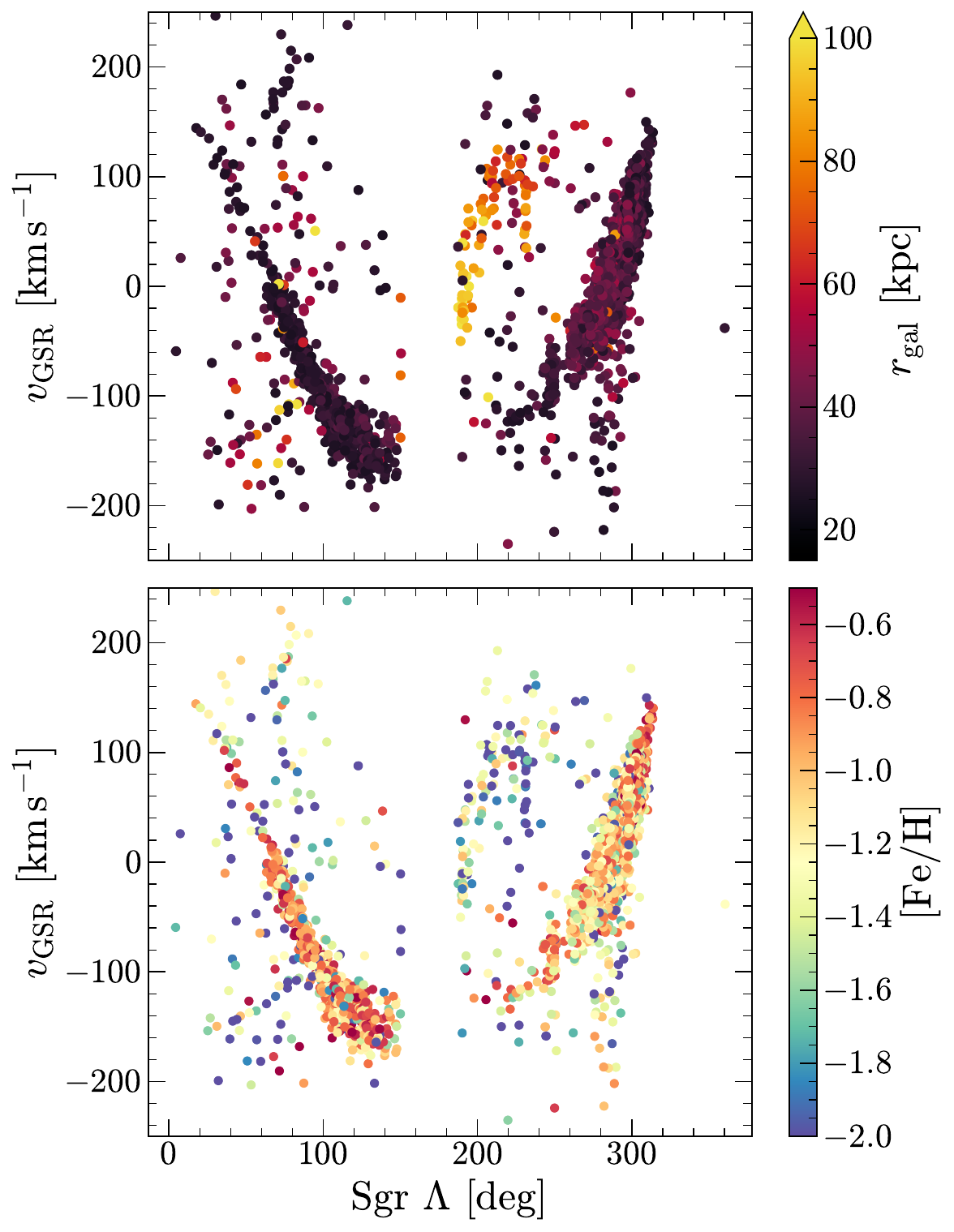}
    \caption{All-sky view of kinematically-selected members of the Sagittarius Stream, showcasing the information in the SDSS-V \minesweepercat{} catalog. Stars in the angular momentum-selected Sgr sample (see Figure~\ref{fig:sgr}) are shown in the space of Galactocentric radial velocity versus longitude along the Sgr stream, colored by Galactocentric distance (top) and metallicity (bottom). 
    }
    \label{fig:sgr_science}
\end{figure}

Figure~\ref{fig:sgr_science} showcases our measurements of Sgr members in the velocity-longitude plane. 
The top panel is colored by distance, and the bottom by metallicity. 
Several interesting features are apparent. 
First, there is a large population of stars that satisfy the kinematic Sgr selection cuts, but lie far from the main Sgr stream's $v_\mathrm{GSR}$ locus. 
Even at comparable distances, these stars are systematically more metal-poor than the main stream body (down to [Fe/H]$~< -2.0$).
This diffuse, metal-poor component of the stream was mapped in H3 Survey \citep{Conroy2019b} data by \cite{Johnson2020a}, who theorized that it might indicate the stellar halo of the Sgr progenitor \citep[see also][]{Gibbons2017}.

The top panel of Figure~\ref{fig:sgr_science} also shows a clump of stars beyond $r_\mathrm{gal} \gtrsim 70$ with markedly different radial velocities compared to the main stream track. 
This is the clearest view to date of the distant `second wrap' of the Sgr stream. 
According to models of the stream, these should be among the earliest-stripped stars in the trailing arm of the stream \citep{Vasiliev2021a, Limberg2023}. 
If the Sgr progenitor galaxy had a spatial metallicity gradient, then this gradient should be imprinted on the stellar stream too. 
This is supported by our data, which show that the distant wrap is systematically metal-poor compared to the closer Sgr arms \citep[bottom panel of Figure~\ref{fig:sgr_science}; see also][]{Limberg2023}.

\begin{figure}
    \centering
    \includegraphics[width=\linewidth]{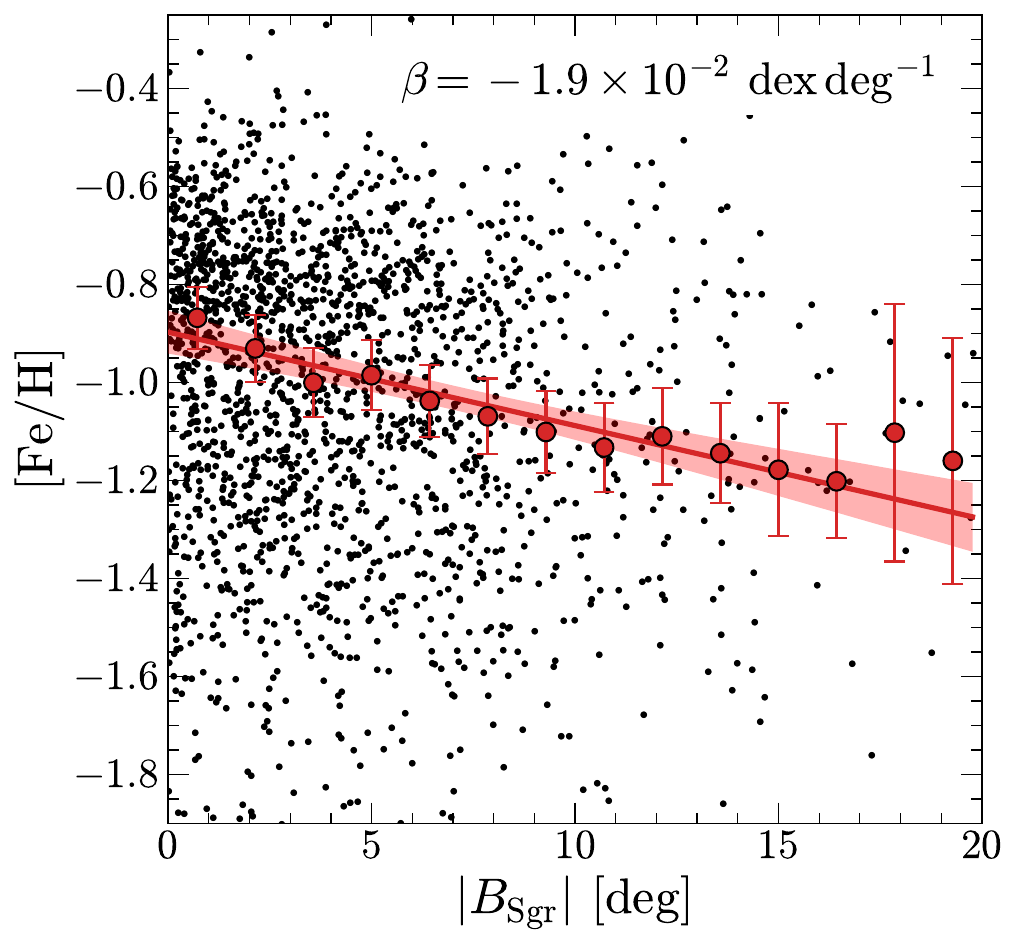}
    \caption{Metallicity of Sgr Stream stars as a function of latitude off the Sgr orbital plane. Stars with $10 < r_\mathrm{gal}/\mathrm{kpc} < 75$ that are within 75~\kms{} of the main Sgr $v_\mathrm{GSR}$ locus (see Figure~\ref{fig:sgr}) are shown. 
    Red markers show the binned average trend, along with a linear fit whose slope $\beta$ is indicated. }
    \label{fig:sgr_gradient}
\end{figure}

The spatial metallicity gradient of the Sgr progenitor should be imprinted not only along the stream, but also across it. 
\cite{Cunningham2024} recently detected the metallicity gradient across the stream (along the stream latitude $B_\mathrm{Sgr}$) for the first time using \textit{Gaia} data, finding a gradient $\approx -1.5\times 10^{-2}\,\mathrm{dex\,deg^{-1}}$. 
Our SDSS-V dataset allows us to investigate this gradient with spectroscopic metallicities for a clean sample of stream stars. 
We select stars along the main stream body of Sgr: within $20^\circ$ of the orbital plane, and within $75$~\kms{} of the stream locus in the $v_\mathrm{GSR}$--$L_\mathrm{Sgr}$ plane. 
This locus was determined by iteratively fitting a 4th-order polynomial to the data shown in Figure~\ref{fig:sgr_science}, and rejecting datapoints $>75$~\kms{} from the polynomial.  
After five iterations, the polynomial closely matches the visual locus, and is used to select main stream members. 
We also restrict the data to $r_\mathrm{gal} < 70$~kpc to remove the distant and metal-poor second wrap of the stream. 

The metallicity gradient across the $B_\mathrm{Sgr}$ for this pure sample of Sgr members is shown in Figure~\ref{fig:sgr_gradient}. 
The binned average (after $3\sigma$ clipping in each bin) is overlaid, and a linear fit is performed on the binned trend. 
We measure a significant gradient of $-1.9\times 10^{-2}\,\mathrm{dex\,deg^{-1}}$
This is consistent with the results from \cite{Cunningham2024}, and based on their N-body simulations, suggests an initial metallicity gradient of $\sim -0.2\,\mathrm{dex\,kpc^{-1}}$ in the Sgr progenitor galaxy.

\begin{figure}
    \centering
    \includegraphics[width=\linewidth]{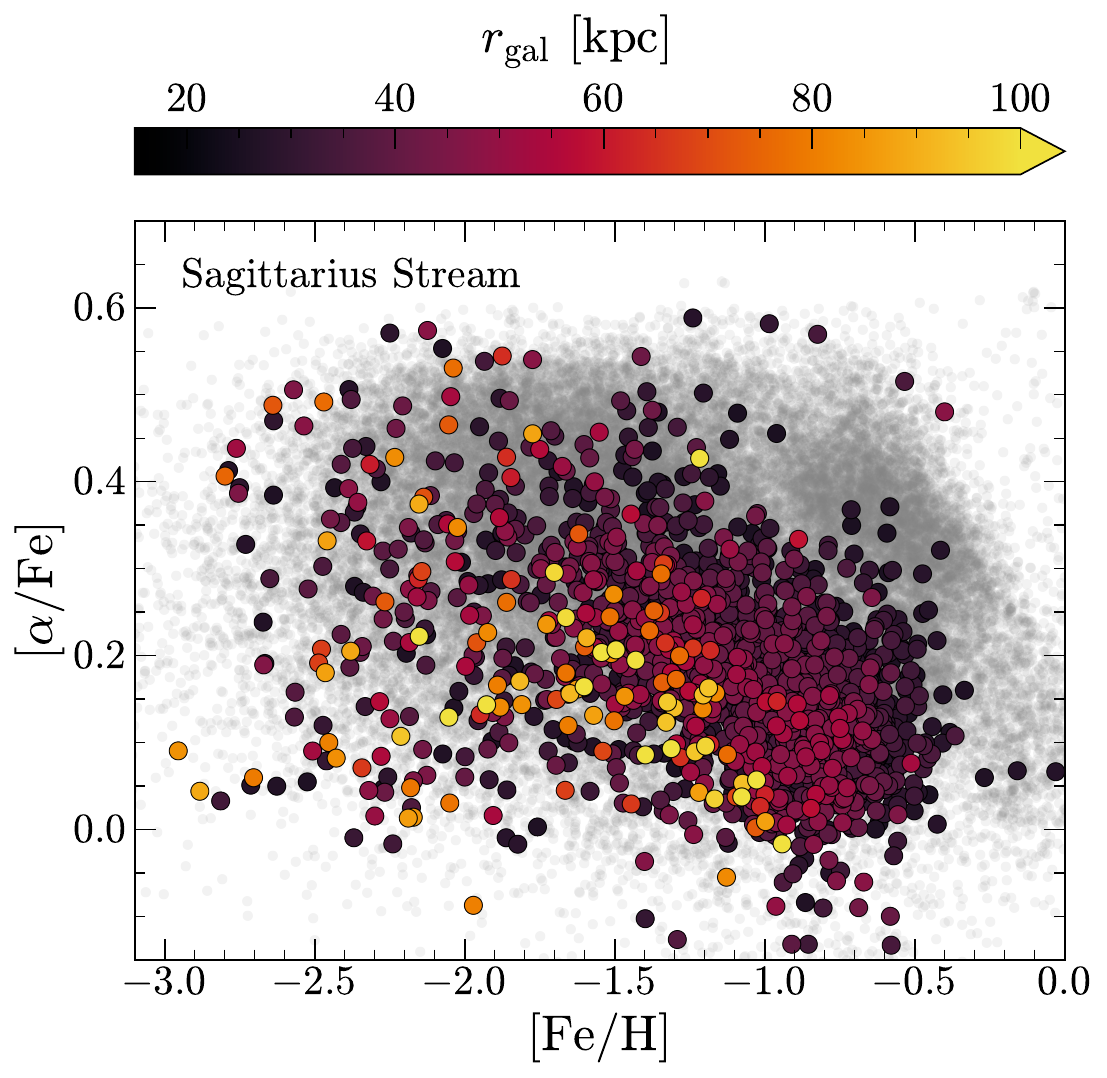}
    \caption{Kinematically-selected Sgr Stream stars in the plane of alpha abundance and metallicity, colored by Galactocentric radius. Points have also been sorted by $r_\mathrm{gal}$ to bring out the more distant structures. Grey points show the distribution of non-Sgr members in the \minesweepercat{} catalog.}
    \label{fig:sgr_tw}
\end{figure}

In addition to the overall metallicity, we also measure the alpha abundance [$\alpha$/Fe], which contains information about what kind of gravitational potential and environment stars were born in. 
For example, stars born in lower-mass dwarf galaxies are known to have lower [$\alpha$/Fe] at fixed [Fe/H] \citep[e.g.,][]{Robertson2005, Hasselquist2021, Cunningham2022}. 
Figure~\ref{fig:sgr_tw} shows the [Fe/H]--[$\alpha$/Fe] plane of kinematically-selected Sgr stars, colored by the Galactocentric distances. 
Not only are the more distant Sgr stars more metal-poor, at fixed metallicity they tend to have lower [$\alpha$/Fe] values than the more nearby debris. 
If the Sgr progenitor had a stellar halo built up from mergers with smaller dwarf galaxies, the natural expectation would be for the earliest-stripped (most distant) Sgr debris to be systematically low-alpha and metal-poor. 
Figure~\ref{fig:sgr_tw} therefore presents potential evidence for an accreted halo in the Sgr progenitor, which should be further investigated in subsequent studies. 

As the survey proceeds, SDSS-V will provide the most complete all-sky inventory of Sgr Stream members out to $100$~kpc and beyond. 
These stars can be used to constrain simulations of the intricate orbital dance between the MW, Sgr, and the Magellanic Clouds \citep[e.g.,][]{Vasiliev2021a}. 
This dataset can also be used to search for the most distant, earliest-stripped debris from Sgr, which should be very sensitive to the progenitor galaxy's initial orbit. 
Past [$\alpha$/Fe] measurements of the stream have been restricted to the most nearby debris \citep[e.g.,][]{Hasselquist2021}. 
The \minesweepercat{} data can trace the Sgr Stream in its entirety, investigating whether the earliest-stripped debris are chemically different from the main stream body. 

\subsection{The Distant Halo}

The furthest halo stars retain a strong dynamical memory of the Milky Way's past merger events, due to their long orbital timescales \citep[e.g.,][]{Majewski2003, Kollmeier2009, Sesar2017a, Belokurov2019a, Chandra2023a, Chandra2023b}. 
Furthermore, distant halo stars allow us to measure the global dynamics of the MW on the largest scales \citep[e.g.,][]{Erkal2021, Bird2021, Han2022b, Petersen2021, Chandra2025}. 
SDSS-V will deliver all-sky spectroscopy of tens of thousands of stars beyond $\gtrsim 10$~kpc, an outer halo sample that is unprecedented in its size and coverage. 

One of the strongest influences in the outer halo is our most dominant neighboring galaxy, the Large Magellanic Cloud (LMC). 
The dark matter halo of the LMC is sufficiently massive that the center of mass of our shared system should meaningfully shift towards the LMC as it falls towards the MW. 
The inner galaxy should rapidly respond to this shift, while the galactic outskirts are left behind --- this induces a `reflex motion' of the outer halo towards the inner galaxy \citep{Weinberg1995, Gomez2015, Garavito-Camargo2019, Erkal2020, Petersen2020}. 
This motion primarily manifests as a radial velocity dipole on the sky, with stars in the southern hemisphere --- where the LMC resides --- being systematically blueshifted, and vice versa in the north. 
The motion should imprint disequilibrium structure throughout the outer halo, and can bias dynamical analyses of the halo if ignored --- for example, inflating the total mass of the MW by up to 50\% \citep{Erkal2020}. 
Only in recent years have observed samples of distant halo stars achieved the depth and sky coverage to detect and measure this reflex motion \citep{Erkal2021, Petersen2021, Yaaqib2024, Chandra2025, Bystrom2024}.

\begin{figure}
    \centering
    \includegraphics[width=\columnwidth]{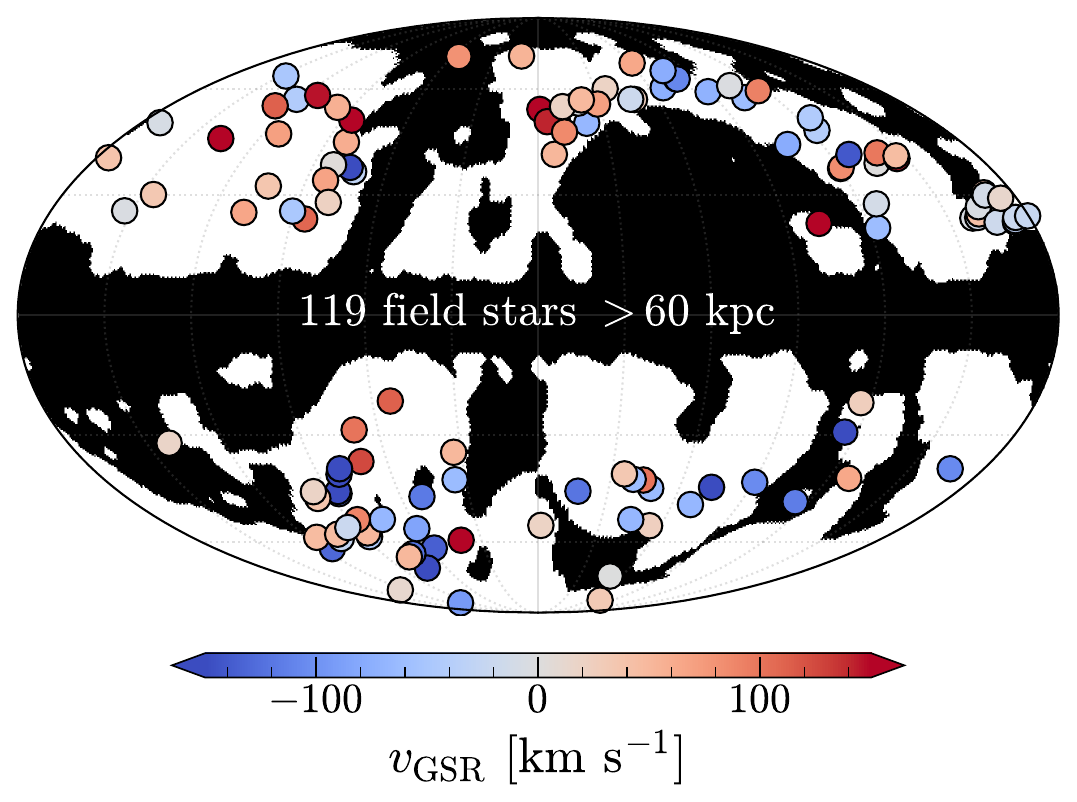}
    \includegraphics[width=\columnwidth]{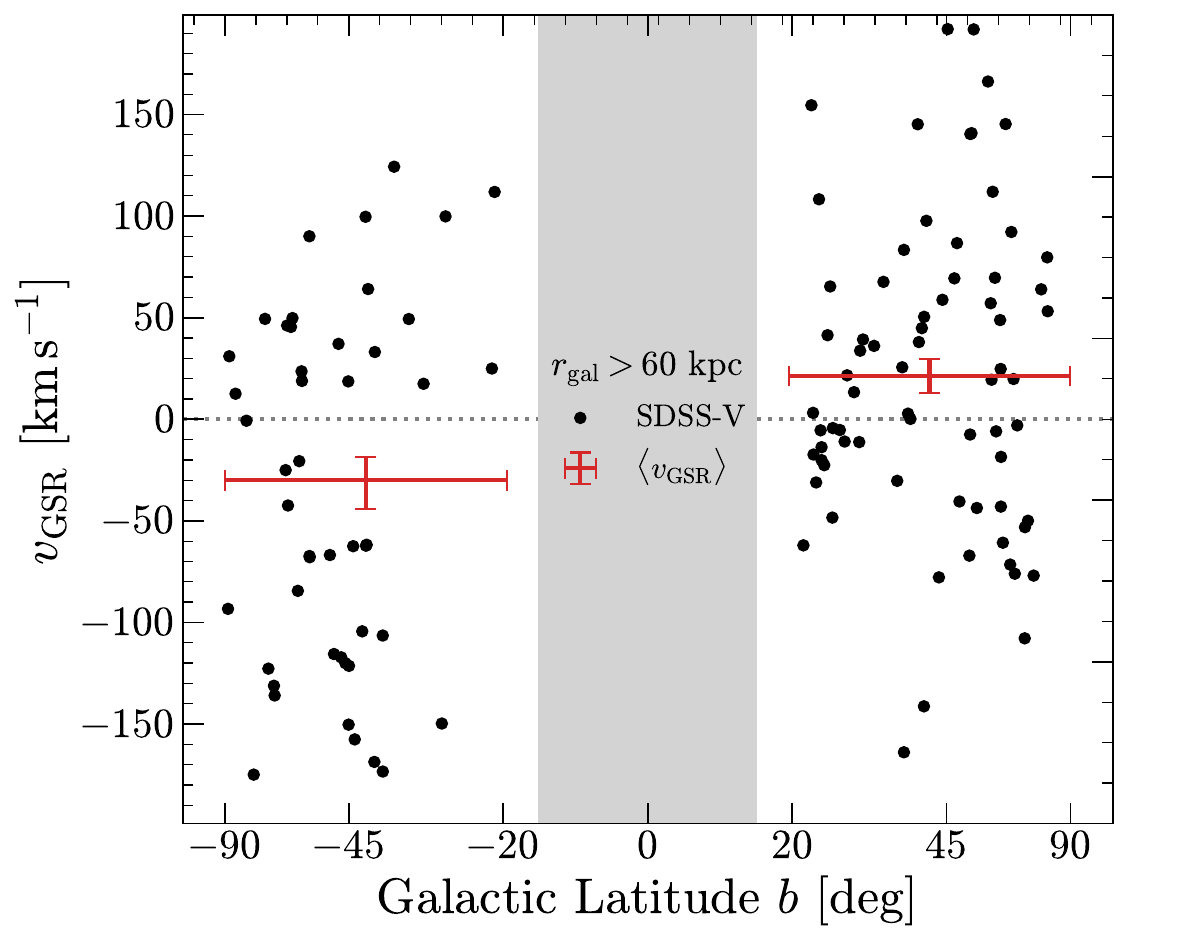}
    \caption{\textbf{Top:} On-sky distribution of field stars with $r_\mathrm{gal} > 60$~kpc in the current catalog, colored by Galactocentric radial velocity. 
    Stars belonging to known substructures like dwarf galaxies, globular clusters, and the Sagittarius Stream have been excised.
    The white background shows the current footprint of the SDSS-V survey, combined with a $|b| > 10^\circ$ selection to remove disk contamination. 
    \textbf{Bottom:} Galactocentric radial velocity of $r_\mathrm{gal} > 60$~kpc stars as a function of (the sine of the) latitude, with the mean value for each hemisphere overlaid, highlighting the North--South asymmetry induced by the LMC. 
    }
    \label{fig:distant}
\end{figure}

SDSS-V is uniquely positioned to measure the reflex motion of the distant halo, due to its all-sky coverage and homogeneous sample selection for luminous red giants. 
Figure~\ref{fig:distant} shows the on-sky distribution of over a hundred field stars already observed by the survey beyond $60$~kpc. 
Known substructures like globular clusters, dwarf galaxies, and stellar streams --- including the Sagittarius 
Stream --- have been removed from the sample, using the \texttt{in\_substructure} flag from the catalog. 
Large on-sky gaps persist in the current survey coverage, which will be filled in by the end of the survey. 
A velocity dipole is already somewhat apparent in this dataset, with stars in the south being on average blueshifted, and vice versa in the north (bottom panel of Figure~\ref{fig:distant}). 

\begin{figure}
    \centering
    \includegraphics[width=\columnwidth]{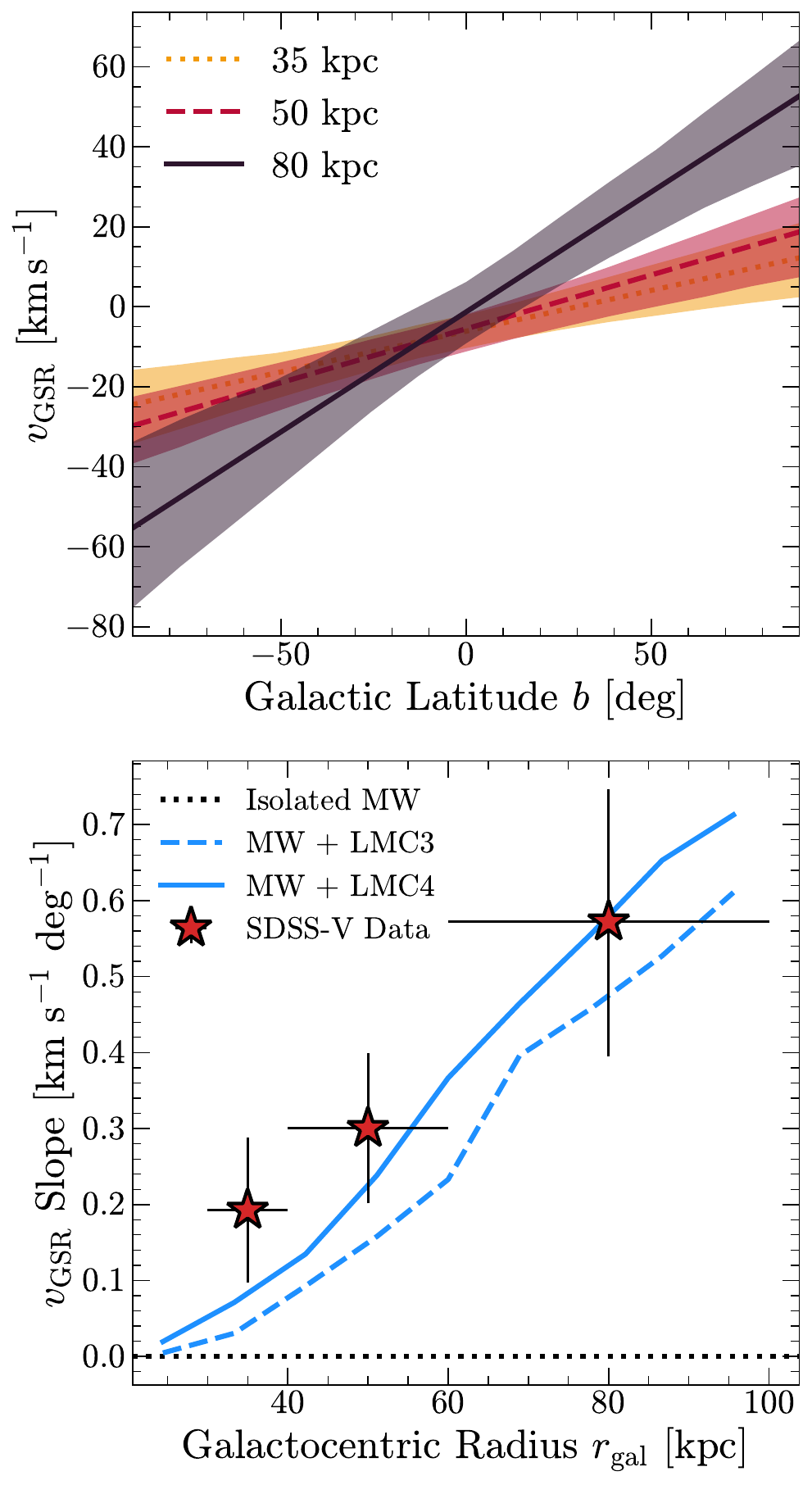}
    \caption{The observed north-south radial velocity dipole of distant halo stars. 
    \textbf{Top:} Linear fits to the run of $v_\mathrm{GSR}$ versus Galactic latitude $b$, with 1$\sigma$ uncertainties shaded, for three radial bins. 
    \textbf{Bottom:} Linear slope of the measurements in the top panel, as a function of Galactocentric radius. 
    In blue, analogous measurements from the simulations of \cite{Garavito-Camargo2019} are shown, with LMC3 (LMC4) corresponding to a total LMC mass of $1.8~(2.5) \times10^{11}\,M_\odot$.}
    \label{fig:vreflex}
\end{figure}

To quantify this dipole further, we investigate the trend of Galactocentric radial velocity as a function of Galactic latitude $b$. 
This is a crude estimator of the reflex-induced dipole ---  the dipole is not predicted to strictly aligned with the Galactic plane, but rather points towards a recent location along the LMC's past orbit \citep{Petersen2021, Chandra2025}. 
However, this trend captures the broad nature of the dipole. 
We fit a linear model to the run of $v_\mathrm{GSR}$ versus $b$, incorporating the velocity uncertainties for each star. 
The resulting fit for three radial bins of Galactocentric distance are shown in the top panel of Figure~\ref{fig:vreflex}. 
The bin edges are $r_\mathrm{gal} = [30,  40,  60, 100]$~kpc. 
A velocity dipole is confidently measured in all bins, with the amplitude of the north-south discrepancy growing as a function of Galactocentric distances. 

The bottom panel of Figure~\ref{fig:vreflex} shows the slope of the linear fit shown in the top panel, as a function of Galactocentric distances for the three bins considered. 
The expectation for a smooth, isotropic halo would be near-zero slope of $v_\mathrm{gsr}$ as a function of latitude $b$. 
For comparison, we repeat the slope measurement using star particles from the simulations of \cite{Garavito-Camargo2019} --- these are N-body simulations of the MW halo's response to the infalling LMC, including the reflex motion. 
Although the uncertainties with the current SDSS data are large, and acknowledging that this linear slope is a crude measure of the velocity dipole, there is good agreement between the data and these models. 
The current data favor a relatively massive LMC that is $\gtrsim 20\%$ the mass of the Milky Way, but no firm conclusion should be drawn from this. 
Future work will fit the SDSS-V data in a full statistical framework to measure the amplitude and direction of the reflex motion \citep[e.g.,][]{Petersen2021, Chandra2025, Brooks2025}.

\section{Conclusions}\label{sec:discuss}

\begin{enumerate}
    \item The SDSS-V halo survey is observing over half a million distant and metal-poor stars across the entire sky. We have described the stellar parameter pipeline and \minesweepercat{} catalog for this survey, which simultaneously compares the spectrum, photometry, and parallax of a star to theoretical models and isochrones in a full Bayesian fitting framework (Figure~\ref{fig:fit}).

    \item \minesweepercat{} stellar parameters are validated with dedicated observations of star clusters (Figure~\ref{fig:cluster_means}), as well as a comparison to the high-resolution APOGEE survey (Figure~\ref{fig:bossms-apogee}). 
    Our parameters perform significantly better for metal-poor stars than the default pipelines, highlighting the importance of our approach for halo stars (Figure~\ref{fig:bossms-bossnet}). 
    Although efforts have been taken to only release a clean subset of reliable parameters, caution must be taken when interpreting prior-driven parameters like stellar ages. 

    \item A key strength of the SDSS-V halo survey is dedicated targeting of the most distant and metal-poor stars in the MW. 
    Using the \minesweepercat{} catalog, we show that the distant giant selection is $\gtrsim 75\%$ efficient at selecting stars beyond $20$~kpc (Figure~\ref{fig:kg}). 
    Furthermore, the \textit{Gaia} XP-based selection to identify very metal-poor stars is $\gtrsim 70\%$ pure, $10\times$ better than a random sample of halo stars (Figure~\ref{fig:mp}). 

    \item We present a broad showcase of the science that is enabled by the \minesweepercat{} catalog. This catalog is already being used to identify metal-poor stars for high-resolution follow-up spectroscopy, leading to the discovery of some of the most chemically peculiar stars yet known in the Milky Way (Figure~\ref{fig:followup}). 

    \item The all-sky nature of SDSS-V enables structures like the Sagittarius Stream to be studied in their entirety, enabling us to map its metallicity structure in 3D (Figures~\ref{fig:sgr}-\ref{fig:sgr_gradient}). 
    We also chart the alpha--abundance of Sgr stars to larger distances than ever before, and present tentative evidence that early--stripped Sgr debris may have belonged to an accreted halo of the Sgr progenitor (Figure~\ref{fig:sgr_tw}). 

    \item SDSS-V is observing one of the largest all-sky samples of field stars in the most distant $r_\mathrm{gal} \gtrsim 50$ outskirts of the MW (Figure~\ref{fig:distant}). 
    These stars trace the MW's dynamics on the largest scales, and already show evidence of the MW's dynamical response to the LMC (Figure~\ref{fig:vreflex}). 

    \item We have re-analyzed over $100,000$ archival spectra from the Sloan Extension for Galactic Understanding and Exploration (SEGUE) survey, resulting in a \segcat{} catalog that is now publicly available (Appendix~\ref{sec:segcat}). 
    
\end{enumerate}

The \minesweepercat{} catalog is publicly released with each data release of SDSS-V\footnote{\url{https://www.sdss.org/dr19/data_access/value-added-catalogs/?vac_id=10007}}. 
The current release is DR19, which only includes data observed from APO up till mid-2023. 
Future releases of the \minesweepercat{} catalog will deliver the most comprehensive, all-sky inventory of Milky Way halo stars, enabling the rich lines of research that have been showcased here. 

\begin{acknowledgments}

We thank 
Andres Almeida, 
Daniel Eisenstein,
David W. Hogg, 
Ben Johnson,
Marina Kounkel,
Sebastian Lepine,
Rohan Naidu,
David Nidever,
and Adrian Price-Whelan
for insightful conversations and feedback.

Funding for the Sloan Digital Sky Survey V has been provided by the Alfred P. Sloan Foundation, the Heising-Simons Foundation, the National Science Foundation, and the Participating Institutions. SDSS acknowledges support and resources from the Center for High-Performance Computing at the University of Utah. SDSS telescopes are located at Apache Point Observatory, funded by the Astrophysical Research Consortium and operated by New Mexico State University, and at Las Campanas Observatory, operated by the Carnegie Institution for Science. The SDSS web site is \url{www.sdss.org}.

SDSS is managed by the Astrophysical Research Consortium for the Participating Institutions of the SDSS Collaboration, including the Carnegie Institution for Science, Chilean National Time Allocation Committee (CNTAC) ratified researchers, Caltech, the Gotham Participation Group, Harvard University, Heidelberg University, The Flatiron Institute, The Johns Hopkins University, L'Ecole polytechnique f\'{e}d\'{e}rale de Lausanne (EPFL), Leibniz-Institut f\"{u}r Astrophysik Potsdam (AIP), Max-Planck-Institut f\"{u}r Astronomie (MPIA Heidelberg), Max-Planck-Institut f\"{u}r Extraterrestrische Physik (MPE), Nanjing University, National Astronomical Observatories of China (NAOC), New Mexico State University, The Ohio State University, Pennsylvania State University, Smithsonian Astrophysical Observatory, Space Telescope Science Institute (STScI), the Stellar Astrophysics Participation Group, Universidad Nacional Aut\'{o}noma de M\'{e}xico, University of Arizona, University of Colorado Boulder, University of Illinois at Urbana-Champaign, University of Toronto, University of Utah, University of Virginia, Yale University, and Yunnan University.

The computations in this paper were run on the FASRC Cannon cluster supported by the FAS Division of Science Research Computing Group at Harvard University. 
This research has made extensive use of NASA's Astrophysics Data System Bibliographic Services.

This work has made use of data from the European Space Agency (ESA) mission {\it Gaia} (\url{https://www.cosmos.esa.int/gaia}), processed by the {\it Gaia} Data Processing and Analysis Consortium (DPAC, \url{https://www.cosmos.esa.int/web/gaia/dpac/consortium}). Funding for the DPAC has been provided by national institutions, in particular the institutions participating in the {\it Gaia} Multilateral Agreement. 

\end{acknowledgments}

\software{\texttt{numpy} \citep{Harris2020}, 
\texttt{scipy} \citep{Virtanen2020}, 
\texttt{matplotlib} \citep{Hunter2007}, 
\texttt{astropy} \citep{AstropyCollaboration2013, AstropyCollaboration2018, AstropyCollaboration2022},
\texttt{gala} \citep{gala,adrian_price_whelan_2020_4159870},
\texttt{MINESweeper} \citep{Cargile2020}
}

\facilities{Gaia, Sloan, PS1, 2MASS, WISE}

\bibliography{library,bib}
\bibliographystyle{aasjournal}

\clearpage

\appendix

\section{Stellar Parameters for SEGUE Stars}\label{sec:segcat}

\begin{figure}
    \centering
    \includegraphics[width=\columnwidth]{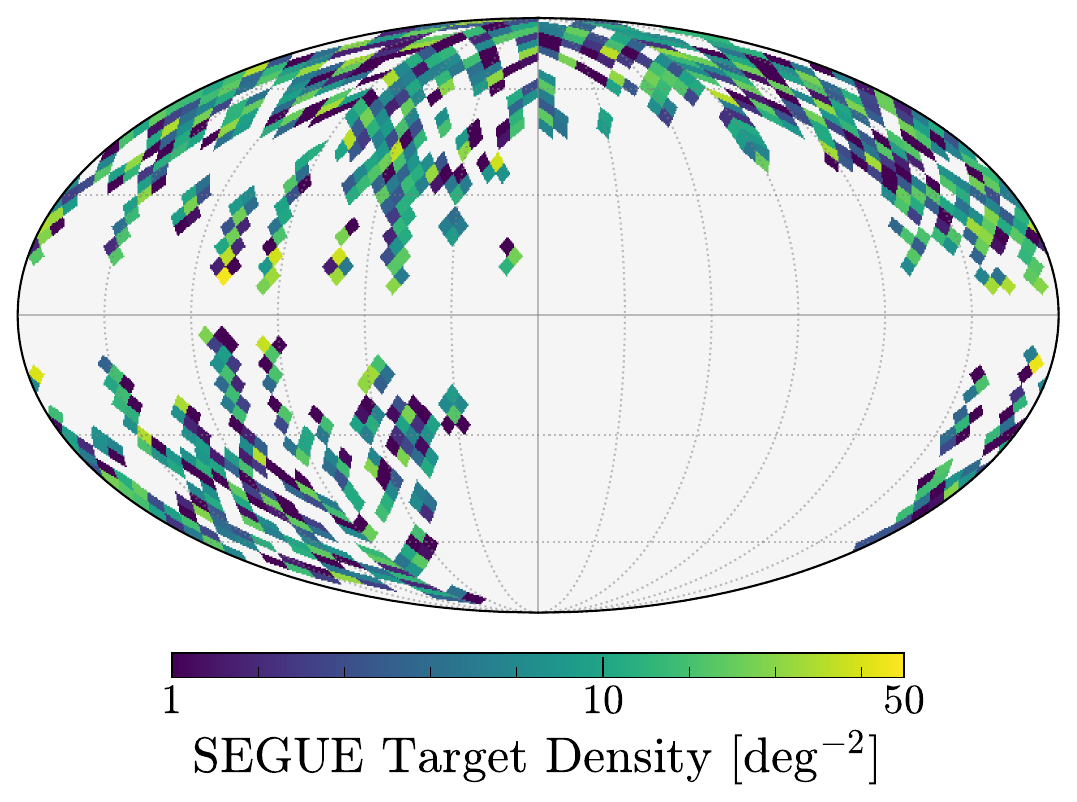}
    \includegraphics[width=\columnwidth]{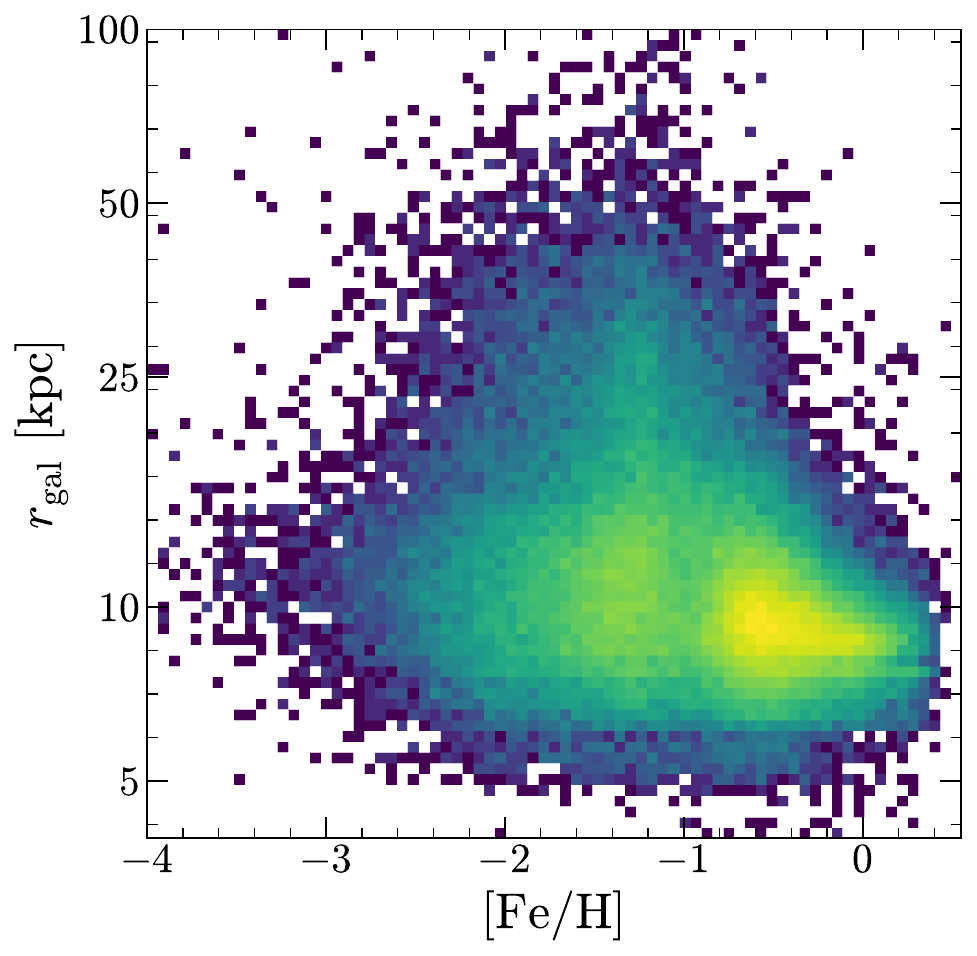}
    \caption{\textbf{Top:} On-sky distribution of stars in the \segcat{} catalog, in Galactic coordinates. 
    Because SEGUE was observed from APO, data only exists for $\delta > -20^\circ$ stars. 
    \textbf{Bottom:} Distribution of stars in the [Fe/H]--$r_\mathrm{gal}$ space, highlighting the halo coverage of the \segcat{} catalog.}
    \label{fig:segcat}
\end{figure}

During the development of \minesweeper{} and the \minesweepercat{} catalog, we also analyzed spectra from the Sloan Extension for Galactic Understanding and Exploration (SEGUE; \citealt{Yanny2009}). 
This was a program that utilized the original SDSS spectrograph at APO to observe over $200,000$ stars from $2003-2009$, drawn from a variety of selections to probe the disk and halo of the MW. 
The top panel of Figure~\ref{fig:segcat} shows the on-sky distribution of SEGUE targets. 

The SEGUE spectra are fit in a manner similar to that described in $\S$\ref{sec:mscode}, utilizing all available photometry and \textit{Gaia} parallaxes in the fit. 
If a single astrophysical source was observed multiple times, we select the highest SNR SEGUE spectrum. 
We apply the same cleanliness cuts as the \minesweepercat{} catalog (described in $\S$\ref{sec:catalog}), resulting in $\approx 100,000$ stars with reliable \minesweeper{} parameters. 
The largest reduction in the sample size comes from the $\log{g} < 4.5$ cut, since SEGUE has a much higher fraction of dwarf stars than the SDSS-V targets (being targeted well before the \textit{Gaia} parallax era). 

The \segcat{} catalog is now publicly released in \cite{Cargile2025}\footnote{\url{https://zenodo.org/records/16105186}}. 
This catalog extends the functionality of SEGUE data by providing a homogeneous set of stellar parameters and isochrone-based distances (bottom panel of Figure~\ref{fig:segcat}). 
As an example of its utility, the \segcat{} catalog was used --- in combination with other \minesweeper{} catalogs --- to map the all-sky dynamics of the MW's outer halo in \cite{Chandra2025}. 

%clearpage

\section{Cluster Validation}\label{sec:clustertab}

Table~\ref{tab:clusters} compares the mean \minesweepercat{} parameters for validation star clusters to values from the literature, and lists references for the literature parameters. 

\begin{deluxetable*}{cccccccc}
\label{tab:clusters}
\tablecaption{Mean \minesweeper{} parameters for calibration clusters, compared to reference values from the literature. 
The uncertainty on the \minesweeper{} parameters is the sigma-clipped standard deviation across all cluster members in SDSS-5.
}
\tablehead{
    \colhead{} & \multicolumn{3}{c}{Literature} & \multicolumn{4}{c}{BOSS-\minesweeper{}} \\
        \colhead{} & \multicolumn{3}{c}{$\overbrace{\hspace{12em}}$} & \multicolumn{4}{c}{$\overbrace{\hspace{18em}}$} \\
    \colhead{Cluster} & \colhead{[Fe/H]} & \colhead{[$\alpha$/Fe]} & \colhead{$d_\mathrm{helio}/\mathrm{kpc}$} & \colhead{$N_\ast$} & \colhead{[Fe/H]} & \colhead{[$\alpha$/Fe]} & \colhead{$d_\mathrm{helio}/\mathrm{kpc}$}
}
\startdata
M4 & $-1.18$ & $0.44$ & $1.9$ & $114$ & $-1.12 \pm 0.11$ & $0.42 \pm 0.09$ & $1.8 \pm 0.1$ \\
M5 & $-1.24$ & $0.23$ & $7.5$ & $86$ & $-1.26 \pm 0.15$ & $0.27 \pm 0.12$ & $6.8 \pm 1.3$ \\
M13 & $-1.50$ & $0.13$ & $7.4$ & $86$ & $-1.57 \pm 0.09$ & $0.28 \pm 0.11$ & $7.8 \pm 1.2$ \\
M55 & $-1.85$ & $0.40$ & $5.3$ & $60$ & $-1.80 \pm 0.10$ & $0.32 \pm 0.10$ & $5.3 \pm 0.9$ \\
M67 & $-0.01$ & $-0.05$ & $0.8$ & $60$ & $-0.09 \pm 0.05$ & $0.07 \pm 0.04$ & $0.8 \pm 0.1$ \\
M80 & $-1.79$ & $0.45$ & $10.3$ & $28$ & $-1.66 \pm 0.14$ & $0.36 \pm 0.10$ & $7.8 \pm 1.2$ \\
M92 & $-2.23$ & $0.14$ & $8.5$ & $44$ & $-2.27 \pm 0.09$ & $0.28 \pm 0.11$ & $8.3 \pm 0.8$ \\
M107 & $-1.01$ & $0.24$ & $5.6$ & $53$ & $-1.06 \pm 0.10$ & $0.42 \pm 0.07$ & $5.5 \pm 1.3$ \\
47 Tuc & $-0.72$ & $0.41$ & $4.2$ & $674$ & $-0.82 \pm 0.07$ & $0.36 \pm 0.08$ & $4.3 \pm 0.5$ \\
NGC 6791 & $0.30$ & $0.08$ & $4.1$ & $27$ & $0.25 \pm 0.10$ & $0.05 \pm 0.05$ & $3.9 \pm 0.4$ \\
NGC 6819 & $0.05$ & $0.00$ & $2.8$ & $58$ & $-0.10 \pm 0.08$ & $0.06 \pm 0.07$ & $2.4 \pm 0.2$ \\
\enddata
\tablecomments{Literature abundance values are mainly taken from \cite{Meszaros2015}, and distances from \cite{Baumgardt2018}. Values are also drawn from \cite[][M4]{Ivans1999}, \cite[][M55]{VandenBerg2018}, \cite[][M67]{Cargile2020}, \cite[][M80]{Carretta2015}, \cite[][Tuc~47]{Roediger2014}, \cite[][NGC~6791]{Boesgaard2015}, and \cite[][NGC~6819]{Meszaros2013}.
}
\end{deluxetable*} %

\section{Data Model}\label{sec:datamodel}

Table~\ref{tab:datamodel} lists all columns in our \minesweepercat{} catalog, along with their units and a brief description. 
A full datamodel is hosted at \url{https://dr19.sdss.org/datamodel/files/MWM_MINESWEEPER/minesweeper.html}. 

\clearpage

\begin{longtable}{lcr}
\caption{Data model for the SDSS-V \minesweepercat{} catalog.} \label{tab:datamodel} \\
\hline
\textbf{Parameter} & \textbf{Unit} & \textbf{Description} \\
\hline
\endfirsthead

\hline
\textbf{Parameter} & \textbf{Unit} & \textbf{Description} \\
\hline
\endhead

\hline
\multicolumn{3}{r}{\emph{Continued on next page}} \\
\hline
\endfoot

\hline
\endlastfoot

\texttt{source\_id}         &                                  & Gaia DR3 Source ID \\
\texttt{G}                  & $\mathrm{mag}$                   & Gaia DR3 G magnitude \\
\texttt{BP}                 & $\mathrm{mag}$                   & Gaia DR3 BP magnitude \\
\texttt{RP}                 & $\mathrm{mag}$                   & Gaia DR3 RP magnitude \\
\texttt{ra}                 & $\mathrm{deg}$                   & Gaia DR3 right ascension \\
\texttt{dec}                & $\mathrm{deg}$                   & Gaia DR3 declination \\
\texttt{parallax}           & $\mathrm{mas}$                   & Gaia DR3 parallax \\
\texttt{parallax\_error}    & $\mathrm{mas}$                   & Gaia DR3 parallax error \\
\texttt{parallax\_over\_error} &                              & Gaia DR3 parallax over error \\
\texttt{pmra}               & $\mathrm{mas\,yr^{-1}}$          & Gaia DR3 proper motion in right ascension \\
\texttt{pmdec}              & $\mathrm{mas\,yr^{-1}}$          & Gaia DR3 proper motion in declination \\
\texttt{pmra\_error}        & $\mathrm{mas\,yr^{-1}}$          & Uncertainty on Gaia DR3 proper motion in right ascension \\
\texttt{pmdec\_error}       & $\mathrm{mas\,yr^{-1}}$          & Uncertainty on Gaia DR3 proper motion in declination \\
\texttt{BP\_RP}             & $\mathrm{mag}$                   & Gaia DR3 BP$-$RP color index \\
\texttt{l}                  & $\mathrm{deg}$                   & Galactic longitude \\
\texttt{b}                  & $\mathrm{deg}$                   & Galactic latitude \\
\texttt{catalogid}          &                                  & SDSS catalog identification number \\
\texttt{sdssid}             &                                  & SDSS object identifier \\
\texttt{field}              &                                  & Field number for the observation \\
\texttt{mjd}                & $\mathrm{day}$                   & Mean modified Julian date of observations \\
\texttt{obs}                &                                  & Three-letter code identifying the observatory (APO or LCO) \\
\texttt{n\_exp}             & $\mathrm{count}$                 & Number of individual 15-min exposures co-added \\
\texttt{n\_spall}           & $\mathrm{count}$                 & Number of spAll rows co-added \\
\texttt{snr}                &                                  & Median signal-to-noise ratio per pixel of the combined spectrum \\
\texttt{ACAT\_ID}           &                                  & Internal identifier for MINESweeper catalog \\
\texttt{EEP}                &                                  & Equivalent evolutionary phase used in isochrone fitting \\
\texttt{EEP\_lerr}         &                                  & Lower uncertainty on EEP \\
\texttt{EEP\_uerr}         &                                  & Upper uncertainty on EEP \\
\texttt{EEP\_err}          &                                  & Typical uncertainty on EEP \\
\texttt{init\_FeH}         & $\mathrm{dex}$                   & Initial stellar metallicity from isochrone fit \\
\texttt{init\_FeH\_lerr}   & $\mathrm{dex}$                   & Lower error on initial [Fe/H] \\
\texttt{init\_FeH\_uerr}   & $\mathrm{dex}$                   & Upper error on initial [Fe/H] \\
\texttt{init\_FeH\_err}    & $\mathrm{dex}$                   & Uncertainty on initial [Fe/H] \\
\texttt{init\_aFe}         & $\mathrm{dex}$                   & Initial stellar alpha-element enhancement from isochrone fit \\
\texttt{init\_aFe\_lerr}   & $\mathrm{dex}$                   & Lower error on initial [\(\alpha\)/Fe] \\
\texttt{init\_aFe\_uerr}   & $\mathrm{dex}$                   & Upper error on initial [\(\alpha\)/Fe] \\
\texttt{init\_aFe\_err}    & $\mathrm{dex}$                   & Uncertainty on initial [\(\alpha\)/Fe] \\
\texttt{init\_Mass}        & $M_\odot$                        & Initial stellar mass from isochrone fit, in solar masses \\
\texttt{init\_Mass\_lerr}  & $M_\odot$                        & Lower uncertainty on initial mass estimate \\
\texttt{init\_Mass\_uerr}  & $M_\odot$                        & Upper uncertainty on initial mass estimate \\
\texttt{init\_Mass\_err}   & $M_\odot$                        & Uncertainty on initial mass estimate \\
\texttt{pc\_0}             &                                  & Continuum polynomial coefficient 0 \\
\texttt{pc\_0\_lerr}       &                                  & Lower error on pc\_0 \\
\texttt{pc\_0\_uerr}       &                                  & Upper error on pc\_0 \\
\texttt{pc\_0\_err}        &                                  & Uncertainty on pc\_0 \\
\texttt{pc\_1}             &                                  & Continuum polynomial coefficient 1 (unitless) \\
\texttt{pc\_1\_lerr}       &                                  & Lower error on pc\_1 \\
\texttt{pc\_1\_uerr}       &                                  & Upper error on pc\_1 \\
\texttt{pc\_1\_err}        &                                  & Uncertainty on pc\_1 \\
\texttt{pc\_2}             &                                  & Continuum polynomial coefficient 2 (unitless) \\
\texttt{pc\_2\_lerr}       &                                  & Lower error on pc\_2 \\
\texttt{pc\_2\_uerr}       &                                  & Upper error on pc\_2 \\
\texttt{pc\_2\_err}        &                                  & Uncertainty on pc\_2 \\
\texttt{pc\_3}             &                                  & Continuum polynomial coefficient 3 (unitless) \\
\texttt{pc\_3\_lerr}       &                                  & Lower error on pc\_3 \\
\texttt{pc\_3\_uerr}       &                                  & Upper error on pc\_3 \\
\texttt{pc\_3\_err}        &                                  & Uncertainty on pc\_3 \\
\texttt{Teff}              & $\mathrm{K}$                     & Effective temperature in Kelvin \\
\texttt{Teff\_lerr}       & $\mathrm{K}$                     & Lower error on effective temperature \\
\texttt{Teff\_uerr}       & $\mathrm{K}$                     & Upper error on effective temperature \\
\texttt{Teff\_err}        & $\mathrm{K}$                     & Uncertainty on effective temperature \\
\texttt{logg}              & $\mathrm{dex}$                   & Logarithm (base 10) of the surface gravity in cm/s² \\
\texttt{logg\_lerr}       & $\mathrm{dex}$                   & Lower error on logg \\
\texttt{logg\_uerr}       & $\mathrm{dex}$                   & Upper error on logg \\
\texttt{logg\_err}        & $\mathrm{dex}$                   & Uncertainty on logg \\
\texttt{logR}              & $\mathrm{dex}$                   & Logarithm of the stellar radius (in solar radii) \\
\texttt{logR\_lerr}       & $\mathrm{dex}$                   & Lower error on logR \\
\texttt{logR\_uerr}       & $\mathrm{dex}$                   & Upper error on logR \\
\texttt{logR\_err}        & $\mathrm{dex}$                   & Uncertainty on logR \\
\texttt{FeH}               & $\mathrm{dex}$                   & Metallicity measurement [Fe/H] \\
\texttt{FeH\_lerr}        & $\mathrm{dex}$                   & Lower error on present [Fe/H] \\
\texttt{FeH\_uerr}        & $\mathrm{dex}$                   & Upper error on present [Fe/H] \\
\texttt{FeH\_err}         & $\mathrm{dex}$                   & Uncertainty on present [Fe/H] \\
\texttt{aFe}               & $\mathrm{dex}$                   & Alpha element enhancement measurement \\
\texttt{aFe\_lerr}        & $\mathrm{dex}$                   & Lower error on alpha enhancement \\
\texttt{aFe\_uerr}        & $\mathrm{dex}$                   & Upper error on alpha enhancement \\
\texttt{aFe\_err}         & $\mathrm{dex}$                   & Uncertainty on alpha enhancement \\
\texttt{Vrad}              & $\mathrm{km\,s^{-1}}$             & Radial velocity in km/s \\
\texttt{Vrad\_lerr}       & $\mathrm{km\,s^{-1}}$             & Lower error on radial velocity \\
\texttt{Vrad\_uerr}       & $\mathrm{km\,s^{-1}}$             & Upper error on radial velocity \\
\texttt{Vrad\_err}        & $\mathrm{km\,s^{-1}}$             & Uncertainty on radial velocity \\
\texttt{Vrot}              & $\mathrm{km\,s^{-1}}$             & Projected rotational velocity. Do not use for science, see discussion of LSF. \\
\texttt{Vrot\_lerr}       & $\mathrm{km\,s^{-1}}$             & Lower error on Vrot \\
\texttt{Vrot\_uerr}       & $\mathrm{km\,s^{-1}}$             & Upper error on Vrot \\
\texttt{Vrot\_err}        & $\mathrm{km\,s^{-1}}$             & Uncertainty on Vrot \\
\texttt{Dist}              & $\mathrm{kpc}$                   & Distance from the Sun in kiloparsecs \\
\texttt{Dist\_lerr}       & $\mathrm{kpc}$                   & Lower error on distance \\
\texttt{Dist\_uerr}       & $\mathrm{kpc}$                   & Upper error on distance \\
\texttt{Dist\_err}        & $\mathrm{kpc}$                   & Uncertainty on distance \\
\texttt{Av}                & $\mathrm{mag}$                   & Visual V-band extinction in magnitudes \\
\texttt{Av\_lerr}         & $\mathrm{mag}$                   & Lower error on $A_V$ \\
\texttt{Av\_uerr}         & $\mathrm{mag}$                   & Upper error on $A_V$ \\
\texttt{Av\_err}          & $\mathrm{mag}$                   & Uncertainty on $A_V$ \\
\texttt{logAge}            & $\log(\mathrm{yr})$              & Logarithm (base 10) of stellar age (years) \\
\texttt{logAge\_lerr}     & $\log(\mathrm{yr})$              & Lower error on log stellar age \\
\texttt{logAge\_uerr}     & $\log(\mathrm{yr})$              & Upper error on log stellar age \\
\texttt{logAge\_err}      & $\log(\mathrm{yr})$              & Uncertainty on log stellar age \\
\texttt{Mass}              & $M_\odot$                        & Current stellar mass in solar masses \\
\texttt{Mass\_lerr}       & $M_\odot$                        & Lower error on current mass estimate \\
\texttt{Mass\_uerr}       & $M_\odot$                        & Upper error on current mass estimate \\
\texttt{Mass\_err}        & $M_\odot$                        & Uncertainty on current mass estimate \\
\texttt{logL}              & $\log(L/L_\odot)$                & Logarithm of stellar luminosity relative to the Sun \\
\texttt{logL\_lerr}       & $\log(L/L_\odot)$                & Lower error on log luminosity \\
\texttt{logL\_uerr}       & $\log(L/L_\odot)$                & Upper error on log luminosity \\
\texttt{logL\_err}        & $\log(L/L_\odot)$                & Uncertainty on log luminosity \\
\texttt{Para}              & $\mathrm{mas}$                   & Fitted parallax \\
\texttt{Para\_lerr}       & $\mathrm{mas}$                   & Lower error on fitted parallax \\
\texttt{Para\_uerr}       & $\mathrm{mas}$                   & Upper error on fitted parallax \\
\texttt{Para\_err}        & $\mathrm{mas}$                   & Uncertainty on fitted parallax \\
\texttt{Age}               & $\mathrm{Gyr}$                   & Stellar age in gigayears. Usually prior-dominated, interpret with caution (see text) \\
\texttt{Age\_lerr}        & $\mathrm{Gyr}$                   & Lower error on age \\
\texttt{Age\_uerr}        & $\mathrm{Gyr}$                   & Upper error on age \\
\texttt{Age\_err}         & $\mathrm{Gyr}$                   & Uncertainty on age \\
\texttt{lnZ}               &                                  & Natural logarithm of the Bayesian evidence \\
\texttt{lnL}               &                                  & Natural logarithm of the likelihood \\
\texttt{lnP}               &                                  & Natural logarithm of the posterior probability from the fit \\
\texttt{chisq\_spec}      &                                  & Chi-square statistic for the spectral fit \\
\texttt{nspecpix}         & $\mathrm{pixels}$                & Number of spectral pixels used in fitting \\
\texttt{chisq\_phot}      &                                  & Chi-square statistic for the photometric fit \\
\texttt{nbands}           & $\mathrm{bands}$                 & Number of photometric bands utilized \\
\texttt{R\_gal}           & $\mathrm{kpc}$                   & Galactocentric radial distance in kiloparsecs \\
\texttt{R\_gal\_err}      & $\mathrm{kpc}$                   & Uncertainty on R\_gal \\
\texttt{X\_gal}           & $\mathrm{kpc}$                   & Galactocentric X-coordinate in kiloparsecs \\
\texttt{X\_gal\_err}      & $\mathrm{kpc}$                   & Uncertainty on X\_gal \\
\texttt{Y\_gal}           & $\mathrm{kpc}$                   & Galactocentric Y-coordinate in kiloparsecs \\
\texttt{Y\_gal\_err}      & $\mathrm{kpc}$                   & Uncertainty on Y\_gal \\
\texttt{Z\_gal}           & $\mathrm{kpc}$                   & Galactocentric Z-coordinate in kiloparsecs \\
\texttt{Z\_gal\_err}      & $\mathrm{kpc}$                   & Uncertainty on Z\_gal \\
\texttt{Vx\_gal}          & $\mathrm{km\,s^{-1}}$             & Galactic Cartesian velocity in X direction \\
\texttt{Vx\_gal\_err}     & $\mathrm{km\,s^{-1}}$             & Uncertainty on Vx\_gal \\
\texttt{Vy\_gal}          & $\mathrm{km\,s^{-1}}$             & Galactic Cartesian velocity in Y direction \\
\texttt{Vy\_gal\_err}     & $\mathrm{km\,s^{-1}}$             & Uncertainty on Vy\_gal \\
\texttt{Vz\_gal}          & $\mathrm{km\,s^{-1}}$             & Galactic Cartesian velocity in Z direction \\
\texttt{Vz\_gal\_err}     & $\mathrm{km\,s^{-1}}$             & Uncertainty on Vz\_gal \\
\texttt{Vr\_gal}          & $\mathrm{km\,s^{-1}}$             & Radial component of galactic velocity \\
\texttt{Vr\_gal\_err}     & $\mathrm{km\,s^{-1}}$             & Uncertainty on Vr\_gal \\
\texttt{Vphi\_gal}        & $\mathrm{km\,s^{-1}}$             & Azimuthal component of galactic velocity \\
\texttt{Vphi\_gal\_err}   & $\mathrm{km\,s^{-1}}$             & Uncertainty on Vphi\_gal \\
\texttt{Vtheta\_gal}      & $\mathrm{km\,s^{-1}}$             & Polar component of galactic velocity \\
\texttt{Vtheta\_gal\_err} & $\mathrm{km\,s^{-1}}$             & Uncertainty on Vtheta\_gal \\
\texttt{V\_tan}           & $\mathrm{km\,s^{-1}}$             & Tangential velocity relative to the Sun \\
\texttt{V\_tan\_err}      & $\mathrm{km\,s^{-1}}$             & Uncertainty on V\_tan \\
\texttt{V\_gsr}           & $\mathrm{km\,s^{-1}}$             & Velocity in the Galactic Standard of Rest frame \\
\texttt{V\_gsr\_err}      & $\mathrm{km\,s^{-1}}$             & Uncertainty on V\_gsr \\
\texttt{Lx}               & $\mathrm{kpc\,km\,s^{-1}}$         & X-component of the angular momentum \\
\texttt{Lx\_err}          & $\mathrm{kpc\,km\,s^{-1}}$         & Uncertainty on Lx \\
\texttt{Ly}               & $\mathrm{kpc\,km\,s^{-1}}$         & Y-component of the angular momentum \\
\texttt{Ly\_err}          & $\mathrm{kpc\,km\,s^{-1}}$         & Uncertainty on Ly \\
\texttt{Lz}               & $\mathrm{kpc\,km\,s^{-1}}$         & Z-component of the angular momentum \\
\texttt{Lz\_err}          & $\mathrm{kpc\,km\,s^{-1}}$         & Uncertainty on Lz \\
\texttt{Ltot}             & $\mathrm{kpc\,km\,s^{-1}}$         & Total angular momentum magnitude \\
\texttt{Ltot\_err}        & $\mathrm{kpc\,km\,s^{-1}}$         & Uncertainty on total angular momentum \\
\texttt{E\_kin\_mw22}     & $\mathrm{km^2\,s^{-2}}$           & Kinetic energy from the MW22 potential model \\
\texttt{E\_kin\_mw22\_err}  & $\mathrm{km^2\,s^{-2}}$         & Uncertainty on kinetic energy from the MW22 model \\
\texttt{E\_pot\_mw22}     & $\mathrm{km^2\,s^{-2}}$           & Potential energy from the MW22 potential model \\
\texttt{E\_pot\_mw22\_err}  & $\mathrm{km^2\,s^{-2}}$         & Uncertainty on potential energy from the MW22 model \\
\texttt{E\_tot\_mw22}     & $\mathrm{km^2\,s^{-2}}$           & Total energy from the MW22 potential model \\
\texttt{E\_tot\_mw22\_err}  & $\mathrm{km^2\,s^{-2}}$         & Uncertainty on total energy from the MW22 model \\
\texttt{ecc\_mw22}         &                                  & Orbital eccentricity from the MW22 model \\
\texttt{ecc\_mw22\_err}    &                                  & Uncertainty on the orbital eccentricity \\
\texttt{R\_apo\_mw22}      & $\mathrm{kpc}$                   & Apocentric radius from MW22 orbit integration \\
\texttt{R\_apo\_mw22\_err} & $\mathrm{kpc}$                   & Uncertainty on the apocenter distance \\
\texttt{R\_peri\_mw22}     & $\mathrm{kpc}$                   & Pericentric radius from MW22 orbit integration \\
\texttt{R\_peri\_mw22\_err}& $\mathrm{kpc}$                   & Uncertainty on the pericenter distance \\
\texttt{z\_max\_mw22}      & $\mathrm{kpc}$                   & Maximum vertical height reached in MW22 orbit integration \\
\texttt{z\_max\_mw22\_err} & $\mathrm{kpc}$                   & Uncertainty on the maximum vertical height \\
\texttt{FLAG}              &                                  & Data quality flag from MINESweeper fit (FLAG==0 selects clean data) \\
\texttt{Sgr\_l}           & $\mathrm{deg}$                   & Sagittarius stream coordinate (longitude) \\
\texttt{Sgr\_b}           & $\mathrm{deg}$                   & Sagittarius stream coordinate (latitude) \\
\texttt{in\_sgr\_L}       &                                  & Flag indicating association with the Sagittarius Stream based on angular momentum \\
\texttt{in\_substructure}  &                                  & Flag indicating membership in a known halo substructure \\
\texttt{in\_cluster}       &                                  & Flag indicating membership in a calibration cluster \\
\texttt{cluster}           &                                  & Identifier for the cluster if membership is determined, otherwise n/a \\
\texttt{kg\_near}         &                                  & Flag indicating membership in $10-30$~kpc K giant target class \\
\texttt{kg\_far}          &                                  & Flag indicating membership in $>30$~kpc K giant target class \\
\texttt{xp\_vmp}          &                                  & Flag indicating membership in very metal-poor (VMP) target class based on Gaia XP \\
\texttt{xp\_mp}           &                                  & Flag indicating membership in metal-poor (MP) target class based on Gaia XP \\
\texttt{bb\_mp}           &                                  & Flag indicating membership in metal-poor (MP) target class based on IR colors \\
\end{longtable}

\hspace{1cm} % force table render

\end{document}